\documentclass[journal]{IEEEtran}
\usepackage{algorithm}
\usepackage{algorithmic}
\usepackage{bbm}

\usepackage{amsmath}
\usepackage{multirow}

\usepackage{graphicx}
\usepackage{epstopdf}
\usepackage{extarrows}
\usepackage{float}
\usepackage[latin1]{inputenc}
\usepackage{tikz}
\usepackage{mathrsfs}
\usepackage{amssymb}
\usepackage{stmaryrd}
\usepackage{mathabx}
\usepackage[normalem]{ulem}
\usepackage{flushend}
\usepackage{amsfonts}
\usetikzlibrary{trees,arrows,automata, positioning, shapes}
\usepackage{tikz-qtree}
\usepackage{multirow}
\usepackage{diagbox}
\hfuzz=\maxdimen
 \usepackage{indentfirst}
\usepackage{subfigure}
\usepackage{array}
\newcolumntype{C}[1]{>{\centering\arraybackslash}p{#1}}
\usepackage[pdftex, bookmarksnumbered, bookmarksopen, colorlinks, citecolor=blue, linkcolor=blue]{hyperref}
\usepackage{pifont}

\newtheorem{definition}{Definition}
\newtheorem{theorem}{Theorem}
\newtheorem{lemma}{Lemma}
\newtheorem{corollary}{Corollary}
\newtheorem{remark}{Remark}
\newtheorem{example}{Example}
\newtheorem{proposition}{Proposition}
\newtheorem{claim}{Claim}

\begin{document}

\title{New Insights into the Decidability of Opacity in Timed Automata}

\author{
    Weilin Deng, Daowen Qiu$^{\star}$, and Jingkai Yang
    \thanks{
    Weilin Deng is with the School of Internet Finance and Information Engineering, Guangdong University of Finance, Guangzhou, 510521, China (e-mail: williamten@163.com).}
    \thanks{Daowen Qiu (Corresponding author) is with the School of Computer Science and Engineering, Sun Yat-Sen University, Guangzhou, 510006, China; and with The Guangdong Key Laboratory of Information Security
Technology, Sun Yat-sen University, Guangzhou 510006, China (e-mail:issqdw@mail.sysu.edu.cn).
}
    \thanks{Jingkai Yang is with the School of Mathematics and Statistics, Yulin Normal University,
    Yulin, 537000, China (e-mail: yangjingkai711@163.com).}
}

\maketitle

\begin{abstract}

This paper investigates the decidability of opacity in timed automata (TA),
a property that has been proven to be undecidable in  general.
First, we address a theoretical gap in recent work by J. An \emph{et al.} (FM 2024) by providing necessary and
sufficient conditions for the decidability of location-based opacity in TA.
Based on these conditions,
we identify a new decidable subclass of TA, called timed automata with integer resets (IRTA),
 where clock resets are restricted to occurring at integer time points.
 We also present a verification algorithm for opacity in IRTA.
 On the other hand, we consider achieving decidable timed opacity
 by weakening the capabilities of intruders.
 Specifically, we show that opacity in general TA becomes decidable under the assumption that
 intruders can only observe time in discrete units.
These results establish theoretical foundations for modeling timed systems
and intruders in security analysis, enabling an effective balance between expressiveness and decidability.
\end{abstract}


\section{Introduction}

Opacity is a crucial security property in discrete event systems (DES) \cite{desbook},
ensuring that secret behaviors of a system remain hidden from external observers (often referred to as intruders in the literature).
Introduced initially in the realm of computer science, opacity has been extensively studied in various models of DES,
including conventional automata (e.g., \cite{opacity-review}-\cite{pre-opacity})
and several extended models (e.g., \cite{probilistic-opacity1}-\cite{petri-opacity1}).
These studies have established foundational results on the verification and enforcement of opacity in DES
without timing constraints.
\par

As many real-world systems involve timing constraints,
the extension of opacity to timed systems has garnered significant attention.
Timed automata (TA) \cite{ad94-timed-automata}
are a powerful model for characterizing systems with real-time constraints.
However,
analyzing opacity in TA remains challenging due to the inherent complexity of their temporal dynamics.
Cassez \cite{timed-opacity} made a seminal contribution by proving that opacity is undecidable for general TA,
even for the subclasses of deterministic TA (DTA) and event-recording automata (ERA) \cite{era1}.
This undecidability result has spurred research in two directions to achieve decidability:
identifying decidable subclasses of TA or weakening the capabilities of intruders reasonably.

\par

In the first direction of research, scholars have identified several decidable subclasses of TA regarding opacity.
 For instance,
 Wang \emph{et al.} \cite{wang2018-realtime-opacity} introduced the first decidable subclass, real-time automata (RTA),
 and  provided an algorithm for verifying initial-state opacity of RTA.
Zhang \cite{Zhang2021-realtime-opacity}-\cite{zhang2024-realtime-opacity} further demonstrated that
state-based opacity verification for RTA is 2-EXPTIME-complete.
Li \emph{et al.} \cite{li2022-realtime-opacity}-\cite{li2025-realtime-opacity} addressed state-based opacity
in constant-time labeled automata (CTLA).
More recently, several studies  \cite{An2024}-\cite{Klein2024} independently verified Cassez's conjecture that opacity
is decidable for discrete-timed automata \cite{timed-opacity}
 by reducing the problem to language inclusion over their tick languages.
Notably, An \emph{et al.} \cite{An2024} proposed a necessary condition and a sufficient condition
for the decidability of opacity in TA, though a complete characterization remains an open problem.

\par
In the second direction of research, scholars have also achieved notable advancements.
For instance,
Andr\'{e} \emph{et al.} \cite{Andre-execution-opacity} investigated opacity under the assumption that  intruders
can only observe the total execution time of a run, without observing individual action timestamps.
Ammar \emph{et al.} \cite{Ammar-bounded-opacity}
studied bounded opacity, a security property requiring  intruders to infer secrets within a predefined time window,
which captures the practical constraint of secrets with temporal validity.

\par
In this paper, we contribute to both directions mentioned above.
On the one hand, we address the open problem posed by An \emph{et al.} \cite{An2024}
by providing a complete characterization of
the conditions under which a TA has decidable location-based opacity.
Specifically, we prove that for a subclass $\mathbb{X}$ of TA,
opacity is decidable if and only if the language inclusion problem for $\epsilon$-$\mathbb{X}$
($\mathbb{X}$ with silent transitions) is decidable.
Based on these conditions, we show that opacity is decidable for timed automata with integer resets (IRTA) \cite{IRTA},
a subclass of TA where clock resets are restricted to occurring at integer time points.
We also provide an algorithm for its verification.
On the other hand,
we consider timed opacity under the scenario where intruders can only observe time in discrete units.
This assumption is reasonable as real-world clocks are always of finite precision.
We establish the notion of \emph{current-location timed opacity against intruders
with discrete-time precision} (CLTO-IDTP) and present its verification algorithm, thereby proving the decidability
of opacity in this scenario.
\par

The rest of this paper is organized as follows. Section II provides the necessary background on TA and opacity.
Section III presents the complete characterization of the conditions for decidable timed opacity,
and discusses the opacity for IRTA.
Section IV addresses the verification of timed opacity against intruders with discrete-time measurement precision.
Finally, Section V concludes the paper.

\section{Preliminaries}

This section introduces the notations and knowledge necessary for the subsequent discussion in this paper.
For more details, see \cite{desbook}, \cite{ad94-timed-automata}, \cite{An2024}.

\par
Let $\mathbb{N}$ be the set of natural numbers, and $[m:n]$ be the set of integers $\{m,m+1,\ldots,n\}$.
Let $\mathbb{R}$ and $\mathbb{R}_{\geq 0}$ be the set of real numbers and  non-negative real numbers, respectively.
We denote by $\lfloor d \rfloor$, $\operatorname{frac}(d)$ and $\lceil d \rceil$
the integer part of $d$, fractional part of $d$, and the smallest integer greater than or equal to $d$, respectively.
Let $|\Omega|$ be the number of elements in the set $\Omega$.
 Let $\Sigma$ be an alphabet, $\Sigma^{*}$ be the set of finite words over $\Sigma$ including the empty word $\epsilon$.
 Let $\Sigma_\epsilon$ be $\Sigma \cup \{\epsilon\}$.
For a valuation $v$ and predicate $\phi$, we write $v \models \phi$ if $v$ satisfies $\phi$.
For a set of valuations $V$, we write $V \models \phi$ if every $v \in V$ satisfies $\phi$.

\par
Finite automata (FA) are a popular model for DES.
A finite automaton is a five-tuple $H = ( \Sigma, Q, Q_0, Q_f, \Gamma )$, where
$\Sigma$ is a finite alphabet,
$Q$ is a finite set of states,
$Q_0 \subseteq Q$ is the set of initial states,
$Q_f \subseteq Q$ is the set of  accepting states,
$\Gamma \subseteq Q \times \Sigma \times Q$ is the transition relation.
Let $\mathfrak{L}_{Q_{1}}^{Q_{2}}({H})$ be the set of all words
that are the labels of paths starting from a state in $Q_{1} \subseteq Q$ and ending at a state in $Q_{2} \subseteq Q$.
The \emph{generated language} and \emph{recognized language} by ${H}$ are
$\mathfrak{L}({H})   = \mathfrak{L}_{Q_{0}}^{Q}({H})$ and
$\mathfrak{L}_f({H}) = \mathfrak{L}_{Q_{0}}^{Q_{f}}({H})$, respectively.

\par

A timed word over  $\Sigma \times \mathbb{R}_{\geq 0}$ is a finite sequence $(\sigma_1, t_1)$ $(\sigma_2, t_2) \cdots (\sigma_n, t_n) \in (\Sigma \times \mathbb{R}_{\geq 0})^*$, $0 \leq t_1 \leq t_2 \leq \cdots \leq t_n$,
where $t_{i}$ is called the \emph{timestamp} of  event $\sigma_{i}$,
indicating that event $\sigma_{i}$ occurs at time $t_{i}$, $i \in[1:n]$.
A \emph{timed language} $\mathcal{L} \subseteq (\Sigma \times \mathbb{R}_{\geq 0})^*$ is a set of timed words.

\par
Timed automata (TA) extend FA by incorporating clocks to model timing constraints,
which are defined as follows.

\par
\begin{definition}
A \emph{timed automaton} is a six-tuple
\begin{align}
  \mathcal{A} = ( \Sigma, L,  L_0,L_f, \mathcal{C}, \Delta ),
\end{align}
where
$\Sigma$ is  a finite alphabet,
$L$ is a finite set of locations,
$L_0 \subseteq L$ is the set of initial locations,
$L_f \subseteq L$ is the set of accepting locations,
$\mathcal{C}$ is a finite set of clocks,
$\Delta \subseteq L \times \Sigma \times \Phi(\mathcal{C}) \times 2^{\mathcal{C}} \times L$ is the transition relation.
Here $\Phi(\mathcal{C})$ is the set of clock constraints of the form
$ g ::= \ c \bowtie k \ | \ g_1 \land g_2 \ | \ true$,
where $c \in \mathcal{C}$, $k \in \mathbb{N}$, and $\bowtie \in \{ <, \leq, =, \geq, > \}$.
Let $\kappa(c)$ be the largest integer constant in the constraints concerning clock $c \in \mathcal{C}$.
If $\epsilon \in \Sigma$, then $\mathcal{A}$ is called an $\epsilon$-TA.
\end{definition}

\par
A \emph{clock valuation} $v: \mathcal{C} \rightarrow \mathbb{R}_{\geq 0}$ assigns a non-negative real value to each clock.  We denote by $\mathbb{R}_{\geq 0}^{\mathcal{C}}$ the set of clock valuations.
For $d \in \mathbb{R}_{\geq 0}$, $(v + d)$ is the valuation where each clock is increased by $d$.
For $r \subseteq \mathcal{C}$,
$[r \mapsto 0]v$  is the valuation where the clocks in $r$ are set to 0 and others remain unchanged.

\par
A \emph{state} of $\mathcal{A}$ is a pair $(l, v)$, where $l \in L$ and $v$ is a clock valuation. A \emph{run} of $\mathcal{A}$ over a timed word
$\omega = (\sigma_1, t_1)(\sigma_2, t_2) \cdots (\sigma_n, t_n)$ is a sequence $(l_0, v_0) \xrightarrow{\tau_1, \sigma_1} (l_1, v_1) \xrightarrow{\tau_2, \sigma_2} \cdots \xrightarrow{\tau_n, \sigma_n} (l_n, v_n)$,
where
$l_0 \in L_0$ and $v_0(c) = 0$ for all $ c \in \mathcal{C}$, and
for each $i \in [1:n] $, there exists $(l_{i-1}, \sigma_i, \phi_i, r_i, l_i) \in \Delta$
such that $(v_{i-1} + \tau_i) \models \phi_i$, $v_i = [r_i \mapsto 0](v_{i-1} + \tau_i)$,
and $\tau_i = t_i - t_{i-1}$ (Let $t_0 = 0$).
The run is \emph{accepting} if $l_n \in L_f$.

\par
For a given TA $\mathcal{A}$,
let $\mathfrak{L}_{L_{1}}^{L_{2}}(\mathcal{A})$ be the set of all timed words
that have a run starting from a location in $L_{1} \subseteq L$ and ending at a location in $L_{2} \subseteq L$.
The \emph{generated language} and \emph{recognized language} by $\mathcal{A}$ are $\mathfrak{L}(\mathcal{A}) = \mathfrak{L}_{L_{0}}^{L}(\mathcal{A})$ and
$\mathfrak{L}_f(\mathcal{A}) = \mathfrak{L}_{L_{0}}^{L_{f}}(\mathcal{A})$, respectively.

\par
Given an observable subset $\Sigma_o \subseteq \Sigma$ of events, the \emph{projection function} $P_{\Sigma_o}$ on timed words with respect to $\Sigma_o$ is a function $P_{\Sigma_o}: (\Sigma \times \mathbb{R}_{\geq 0})^* \rightarrow (\Sigma_o \times \mathbb{R}_{\geq 0})^*$ defined recursively as follows: $P_{\Sigma_o}(\epsilon) = \epsilon$, and
\begin{align}\label{eq:observation}
P_{\Sigma_o}((\sigma, t) \cdot \omega) = \begin{cases}
(\sigma, t) \cdot P_{\Sigma_o}(\omega) & \text{if } \sigma \in \Sigma_o, \\
P_{\Sigma_o}(\omega) & \text{otherwise}.
\end{cases}
\end{align}
 $P_{\Sigma_o}$ can be extended to timed languages as follows:
for timed language $\mathcal{L}$,
$P_{\Sigma_o}(\mathcal{L}) = \{ P_{\Sigma_o}(\omega) \mid \omega \in \mathcal{L} \}$.

\par
To address the infinite state space of TA, region equivalence is employed to partition clock valuations into finitely many equivalence classes.

\par
\begin{definition}
Given a TA $\mathcal{A}= ( \Sigma, L, L_0, L_f, \mathcal{C}, \Delta )$ with the largest
 constant $\kappa(c)$, $c \in \mathcal{C}$.
Two clock valuations $v_1, v_2$ are \emph{equivalent}, denoted $v_1 \cong v_2$, if
\begin{itemize}
  \item
 for all $c \in \mathcal{C}$, either $\lfloor v_1(c) \rfloor = \lfloor v_2(c) \rfloor$, or both $v_1(c) > {\kappa}(c)$ and $v_2(c) > {\kappa}(c)$,
  \item
 for all $c \in \mathcal{C}$, if $v_1(c) \leq {\kappa}(c)$, then $\operatorname{frac}(v_1(c)) = 0$ if and only if $\operatorname{frac}(v_2(c)) = 0$,
  \item
 for all $c_1, c_2 \in \mathcal{C}$, if $v_1(c_1) \leq {\kappa}(c_1)$ and $v_1(c_2) \leq {\kappa}(c_2)$,
  then $\operatorname{frac}(v_1(c_1)) \leq \operatorname{frac}(v_1(c_2))$ if and only if $\operatorname{frac}(v_2(c_1)) \leq \operatorname{frac}(v_2(c_2))$,
\end{itemize}
\end{definition}

\par
The relation $\cong$ is an equivalence relation of finite index, and it naturally induces a
finite partition of $\mathbb{R}_{\geq 0}^{\mathcal{C}}$:
$Reg(\mathcal{A}) =  \{ [v]  | v \in \mathbb{R}_{\geq 0}^{\mathcal{C}}  \}$ where
$[v] = \{v^{'} \in \mathbb{R}_{\leq 0}^{\mathcal{C}} | v \cong v^{'}\}$ is called a \emph{region}.
Let ${IReg}(\mathcal{A})$ denote the set of regions containing only integer valuations,
i.e., ${IReg}(\mathcal{A}) = \{[v] \mid \forall c \in \mathcal{C}, v(c) \in \{0, 1, \ldots, \kappa(c) + 1\}\}$.
We denote $[\mathbf{0}]$ by the valuation that assigns 0 to all clocks.
 The partial order $\preceq$ on regions is defined as follows:
$R \preceq R'$, if for any $v \in R$, there exists a $d_{v} \in \mathbb{R}_{\geq 0}$ such that $(v+d_{v}) \in R'$.
\par

\begin{definition} \label{def:ra}
Given a TA $\mathcal{A} = ( \Sigma, L, L_0, L_f, \mathcal{C}, \Delta )$,
the corresponding \emph{region automaton} $\mathcal{R}(\mathcal{A})$ is the FA
$
    \mathcal{R}(\mathcal{A}) = (\Sigma, Q, Q_0, Q_f, \Gamma),
$
where
     $Q = L \times Reg(\mathcal{A})$ is the finite set of states,
  $Q_0 = L_0 \times [\mathbf{0}]$ is the set of initial states,
  $Q_f = L_f \times Reg(\mathcal{A})$ is the set of accepting states, and
 $\Gamma \subseteq Q \times \Sigma \times Q$ is the transition relation.
    A transition $((l, R), \sigma, (l', R')) \in \Gamma$ if and only if there exists a region $R^{''}$ and
    $(l, \sigma, \phi, r, l') \in \Delta$ such that
         $R \preceq R^{''}$, $R^{''} \models \phi$, and $R' = \{ [r \mapsto 0]v \mid v \in R^{''} \}$.
\end{definition}

The region automaton is an abstraction of a TA into an FA
which represents the (untimed) behaviors of the TA,
that is, $\operatorname{Untime}(\mathfrak{L}(\mathcal{A})) = \mathfrak{L}(\mathcal{R}(\mathcal{A}))$.

\par

\begin{definition} \label{def:integralautomaton}
Given a TA $\mathcal{A} = ( \Sigma, L, L_0, L_f, \mathcal{C}, \Delta )$,
the corresponding \emph{integral automaton} is the FA
$\mathcal{A}^\checkmark = (\Sigma \cup \{\checkmark\}, Q^\checkmark, Q_0^\checkmark, Q_f^\checkmark, \Delta^\checkmark)$,
where
 $Q^\checkmark = L \times IReg(\mathcal{A})$ is the set of states,
 $Q_0^\checkmark = L_0 \times \{[\mathbf{0}]\}$ is the set of initial states,
 $Q_f^\checkmark = L_f \times IReg(\mathcal{A})$ is the set of accepting states, and
 $\Delta^\checkmark \subseteq Q^\checkmark \times (\Sigma \cup \{\checkmark\}) \times Q^\checkmark$ is the
     transition relation, including two types of transitions:
    \begin{itemize}
        \item action-transition: $(l, [v]) \xrightarrow{\sigma} (l', [v'])$, if there exists
        $(l, \sigma, \phi, r, l') \in \Delta$ in $\mathcal{A}$, and $[v], [v'] \in IReg(\mathcal{A})$ such that $ [v] \models \phi$, and $v' = [r \mapsto 0]v$.
        \item tick-transition: $(l, [v]) \xrightarrow{\checkmark} (l, [v'])$, if there exists $[v], [v'] \in IReg(\mathcal{A})$ such that $[v'] = [v + 1]$.
    \end{itemize}

The integral automaton $\mathcal{A}^\checkmark$ simulates the behavior of $\mathcal{A}$ under discrete-time semantics, where time progresses in integer units, and actions are triggered only at integer clock valuations.
\end{definition}

Given a TA $\mathcal{A}$, an observable event set $\Sigma_o \subseteq \Sigma$,
a set of secret locations $L_s \subseteq L$,
and a set of non-secret locations $L_{ns} \subseteq L$ (it is \emph{not} necessarily  $L_{ns}\cup L_{s} = L$
in the general case),
the notion of \emph{Current-Location Timed Opacity (CLTO)} is defined as follows.
\begin{definition} \label{def:clto}
Let $\mathcal{A}$ be a TA.
$\mathcal{A}$ is \emph{current-location timed opaque} with respect to $\Sigma_o$, $L_s$, and $L_{ns}$,
if $\forall \omega \in \mathfrak{L}_{L_{0}}^{L_{s}}(\mathcal{A})$,
\begin{align}\label{eq:clto}
 \exists \omega' \in \mathfrak{L}_{L_{0}}^{L_{ns}}(\mathcal{A}) \text{ such that }
          P_{\Sigma_o}(\omega) = P_{\Sigma_o}(\omega'),
\end{align}
or equivalently $P_{\Sigma_o}(\mathfrak{L}_{L_{0}}^{L_{s}}(\mathcal{A})) \subseteq
P_{\Sigma_o}(\mathfrak{L}_{L_{0}}^{L_{ns}}(\mathcal{A}))$.
\end{definition}

\par
The \emph{CLTO} of TA shares a similar intuitive meaning with
the current-state opacity (CSO) of FA (for the formal definition of CSO, see \cite{ifo}).
Intuitively, these two notions require that for each secret behavior,
there exists a non-secret behavior such that both produce identical observations to intruders.

\section{A Complete Characterization for Decidable Timed Opacity}

In this section, we establish necessary and sufficient conditions for the decidability of CLTO in TA.
Building upon these theoretical foundations, we further develop a verification algorithm for CLTO in IRTA.

\subsection{Sufficient and Necessary Conditions for Decidable CLTO}

\par
The CLTO problem for a subclass $\mathbb{X}$ of TA is defined as follows. Given an $\mathbb{X}$ automaton $\mathcal{A}$,
answer whether or not $\mathcal{A}$ satisfies CLTO. If there exists an algorithm to solve the CLTO problem for $\mathbb{X}$,
it is said that the CLTO problem for  $\mathbb{X}$ is decidable.

\par
 A necessary condition 
and a sufficient condition for the decidability of CLTO in TA are presented in \cite{An2024}, which are as follows.

\begin{proposition}
If the CLTO problem for a subclass $\mathbb{X}$ of TA is decidable,
then the universality problem for $\mathbb{X}$ is decidable.
\end{proposition}

\begin{proposition}
If a subclass $\mathbb{X}$ of TA is closed under product, complementation, and projection,
then the CLTO problem for $\mathbb{X}$ is decidable\footnote{Actually, in \cite{An2024}, this sufficient condition is presented for the language-based timed opacity (LBTO). However, it also holds for CLTO, as the CLTO problem can be reduced to the LBTO problem for general TA.}.
\end{proposition}

\par
First of all, we assert that the  necessary condition is not sufficient,
 and the sufficient condition is not necessary.

 \par

\begin{claim}
There exists a subclass $\mathbb{X}$ of TA for which the universality problem is decidable,
yet the CLTO problem is undecidable.
\end{claim}
\begin{IEEEproof}
  We show that deterministic timed automata (DTA) are such a subclass of TA.
  First, the universality problem for DTA is decidable,
  which has been proved in \cite{ad94-timed-automata}.
  Second, the LBTO problem for DTA is undecidable \cite{timed-opacity},
  and the LBTO problem and the CLTO problem for DTA are inter-reducible \cite{An2024}.
  Hence, the CLTO problem for DTA is undecidable.
\end{IEEEproof}

\par

\begin{claim}
There exists a subclass $\mathbb{X}$ of TA for which
 the CLTO problem is decidable, yet
 $\mathbb{X}$ is \emph{not} closed under certain operations among product, complementation, and projection.
\end{claim}
\begin{IEEEproof}
We show that the timed automata with integer resets (IRTA) are such a subclass of TA.
First, the CLTO problem for IRTA is decidable (we will show it in the next subsection).
Second, IRTA has been shown to be strictly less expressive than $\epsilon$-IRTA in \cite{IRTA},
which implies that IRTA is \emph{not} closed under the operation of projection.
\end{IEEEproof}

 \par
To address the open problem posed by J. An et al. \cite{An2024} concerning a complete characterization
of decidable timed opacity, we present the necessary and sufficient conditions for its decidability as follows.

 \par

\begin{theorem} \label{th:CLTO-ns}
  The CLTO problem for a subclass $\mathbb{X}$ of TA is decidable
  if and only if the inclusion problem for $\epsilon$-$\mathbb{X}$
  (i.e., $\mathbb{X}$ with silent transitions) is decidable.
\end{theorem}
\begin{IEEEproof}
  It is sufficient to show that the CLTO problem for $\mathbb{X}$ and
  the inclusion problem for $\epsilon$-$\mathbb{X}$ are inter-reducible.
  \par
  First, we show that the CLTO problem for $\mathbb{X}$ can be reduced to
    the inclusion problem for $\epsilon$-$\mathbb{X}$.
    Given an $\mathbb{X}$ automaton
    $\mathcal{A} = ( \Sigma, L, L_0, L_f, \mathcal{C}, \Delta )$ with
    an observable event set $\Sigma_o \subseteq \Sigma$,
    a set of secret locations $L_s \subseteq L$,
    and a set of non-secret locations $L_{ns} \subseteq L$,
    construct two $\epsilon$-$\mathbb{X}$ $\mathcal{B}$ and $\mathcal{C}$ from $\mathcal{A}$ as follows:
    $\mathcal{B} = ( \Sigma^{B}, L^{B}, L_0^{B}, L_f^{B}, \mathcal{C}^{B}, \Delta^{B} )$ and
    $\mathcal{C} = ( \Sigma^{C}, L^{C}, L_0^{C}, L_f^{C}, \mathcal{C}^{C}, \Delta^{C} )$,
    where $\Sigma^{B} = \Sigma^{C} = \Sigma_{o} \cup \{\epsilon\}$, $L^{B}=L^{C}=L$, $L_0^{B}=L_0^{C}=L_0$,
    $L_f^{B} = L_{s}$, $L_f^{C} = L_{ns}$, $\mathcal{C}^{B} = \mathcal{C}^{C} = \mathcal{C}$,
    and
    $\Delta^{B} = \Delta^{C} = \{ (l, \epsilon, \phi, r, l') | \exists (
               \sigma \in \Sigma \backslash \Sigma_{o} \wedge (l, \sigma, \phi, r, l') \in \Delta\}) \}      \cup
     \{ (l, \sigma, \phi, r, l') | \exists  (\sigma \in \Sigma_{o} \wedge (l, \sigma, \phi, r, l') \in \Delta) \} $.
    That is, $\mathcal{B}$ and $\mathcal{C}$ are obtained from $\mathcal{A}$ by replacing the unobserved events with
     $\epsilon$ in transitions and by setting the set of accepting locations to $L_{s}$ and $L_{ns}$, respectively.
    By the definitions of  projection and timed languages, we have
    $P_{\Sigma_{o}}(\mathfrak{L}_{L_{0}}^{L_{s}}(\mathcal{A})) = \mathfrak{L}_{f}(\mathcal{B})$ and
    $P_{\Sigma_{o}}(\mathfrak{L}_{L_{0}}^{ L_{ns}}(\mathcal{A})) = \mathfrak{L}_{f}(\mathcal{C})$.
    By Definition \ref{def:clto}, $\mathcal{A}$ satisfies CLTO if and only if
    $\mathfrak{L}_{f}(\mathcal{B}) \subseteq \mathfrak{L}_{f}(\mathcal{C})$.

    \par
    Second, we show that the inclusion problem for $\epsilon$-$\mathbb{X}$ can be reduced to
    the CLTO problem for $\mathbb{X}$. Consider any two $\epsilon$-$\mathbb{X}$ automata
    $\mathcal{B} = ( \Sigma^{B}, L^{B}, L_0^{B}, L_f^{B}, \mathcal{C}^{B}, \Delta^{B} )$ and
    $\mathcal{C} = ( \Sigma^{C}, L^{C}, L_0^{C}, L_f^{C}, \mathcal{C}^{C}, \Delta^{C} )$.
     Without loss of generality,
     assume both the locations and clocks in $\mathcal{B}$ are disjoint from those in $\mathcal{C}$,
     since any overlapping elements can be distinguished via a renaming procedure.
    Construct two $\mathbb{X}$ automata $\mathcal{B'}$ and $\mathcal{C'}$ from $\mathcal{B}$ and $\mathcal{C}$,
    respectively, by replacing $\epsilon$ with a new event $\sigma^{+}$ in transition relations,
    and updating the alphabets by removing the $\epsilon$ and adding the $\sigma^{+}$.
    Define the $\mathbb{X}$ automaton $\mathcal{A}$ as follows:
    $\mathcal{A} = (\mathcal{B'} \cup \mathcal{C'}) = ( \Sigma^{A}, L^{A}, L_0^{A}, L_f^{A}, \mathcal{C}^{A}, \Delta^{A})$.
    Let $L_{s} = L_f^{B}$, $L_{ns} = L_f^{C}$,
     and $\Sigma_{o} = \Sigma^{A} \setminus \{ \sigma^{+} \}$.
    Based upon the aforementioned settings, we have
    $ P_{\Sigma_{o}} (\mathfrak{L}_{L_{0}^{{A}}}^{L_{s}} (\mathcal{A})) = \mathfrak{L}_{f}(\mathcal{B})$
    and
    $P_{\Sigma_{o}} (\mathfrak{L}_{L_{0}^{{A}}}^{L_{ns}}(\mathcal{A})) = \mathfrak{L}_{f}(\mathcal{C})$.
    Therefore,
    $\mathfrak{L}_{f}(\mathcal{B}) \subseteq \mathfrak{L}_{f}(\mathcal{C})$ is equivalent to
    $P_{\Sigma_{o}} (\mathfrak{L}_{L_{0}^{{A}}}^{L_{s}}(\mathcal{A})) \subseteq
    P_{\Sigma_{o}} (\mathfrak{L}_{L_{0}^{{A}}}^{L_{ns}}(\mathcal{A}))$, indicating that
    $\mathcal{A}$ satisfies CLTO by Definition \ref{def:clto}.
\end{IEEEproof}

\subsection{CLTO Verification in IRTA}

Timed automata with integer resets (IRTA), introduced by  P. V. Suman \emph{et al.} \cite{IRTA},
 are a subclass of TA where clock resets are restricted to occurring at integer time points.
 Formally, an IRTA is defined as follows.
 \begin{definition}
\label{def:irta}
A \emph{timed automaton with integer resets} is a TA
$
\mathcal{A} = ( \Sigma, L, L_0, L_f, \mathcal{C}, \Delta )
$
in which for every transition $(l, \sigma, \phi, r, l') \in \Delta$,
if $r \neq \emptyset$, then $\phi$ contains at least one atomic constraint $c=k$, in which $c \in \mathcal{C}$ and $k \in \mathbb{N}$.
\end{definition}

\par
IRTA naturally model distributed real-time systems and specifications in business processes or web services \cite{IRTA}.
For instance, IRTA can capture constraints like ``a transaction must be completed within three days"
or periodic behaviors in control systems with multiple electronic control units.

\par
IRTA have a perfect property that they do \emph{not} distinguish between the time stamps of events occurring
within a unit open interval $(i,i+1)$.
The following augmented TA extends timed words of $\mathcal{A}$ by inserting events $\checkmark$ and $\delta$ to
record the global integer time points and fraction time points  \cite{IRTA}.

\par

\begin{definition}\label{def:ma}
Given an IRTA $\mathcal{A} = ( \Sigma, L, L_0, L_f, \mathcal{C}, \Delta )$,
the augmented TA $\mathcal{\overline{A}}$ is defined as follows:
$
\mathcal{\overline{A}} = (\Sigma \cup \{ \delta, \checkmark \}, L', L_0', L_f', \mathcal{C} \cup \{c\}, \Delta')
$
where
\begin{itemize}
    \item $L' = \{l^0, l^+ \mid l \in L\}$ splits each location into its \emph{integral time phase} ($l^0$) and
    \emph{fractional time phase} ($l^+$),
    \item $L_0' = \{l^0 \mid l \in L_0\}$ initializes all clocks to integral phases,
    \item A fresh clock $c \notin \mathcal{C}$ tracks the updating of time phases,
    \item Transitions $\Delta'$ are defined as:
    \begin{align*}
     \Delta' = &\ \{ (l^0_1, \sigma, \phi \land c=0, r, l^0_2) \mid (l_1, \sigma, \phi, r, l_2) \in \Delta  \} \\
            &\ \{ (l^+_1, \sigma, \phi \land 0<c<1, r, l^+_2) \mid (l_1, \sigma, \phi, r, l_2) \in \Delta   \} \\
        \cup &\ \{ (l^0, \delta,     0<c<1, \emptyset, l^+) \mid l \in L \}  \\
        \cup &\ \{ (l^+, \checkmark, c=1,   \{c\},     l^0) \mid l \in L \}.
    \end{align*}
    \item Accepting states $L_f' = \{l^0, l^+ \mid l \in L_f\}$.
\end{itemize}
\end{definition}

\par
It was proved in \cite{IRTA} that the language inclusion problem for $\epsilon$-IRTA
can be reduced to the language inclusion problem for the region automata of the augmented TA.
\begin{lemma} \label{lemma:1}
  Let $\mathcal{A}$ and $\mathcal{B}$ be two $\epsilon$-IRTA.
  $ \mathfrak{L}_f (\mathcal{A}) \not\subseteq \mathfrak{L}_f(\mathcal{B}) $ if and only if
  $ \mathfrak{L}_f(\mathcal{R(\overline{A})} ) \not\subseteq \mathfrak{L}_f(\mathcal{R(\overline{B})} )$.
\end{lemma}

\par
Note that the region automata are conventional automata and their inclusion problem is decidable.
Therefore, the inclusion problem for $\epsilon$-IRTA is decidable.
By Theorem \ref{th:CLTO-ns}, the CLTO problem for IRTA is also decidable.

\par
In the following, we present a verification algorithm for CLTO in an IRTA (Algorithm \ref{alg:clto}).
\par

\begin{algorithm}[!h]
\caption{Verification of Current-Location Timed Opacity (CLTO) for Timed Automata with Integer Resets (IRTA)}
\label{alg:clto}
\textbf{Input}: An IRTA $\mathcal{A} = (\Sigma, L, L_0, L_f, \mathcal{C}, \Delta)$,
a set of observable events $\Sigma_o \subseteq \Sigma$,
a set of secret locations $L_s \subseteq L$, and a set of non-secret locations $L_{ns} \subseteq L$.

\textbf{Output}: ``YES" if $\mathcal{A}$ satisfies CLTO; 
                 ``NO" otherwise.

\begin{algorithmic}[1]

\STATE Obtain $\mathcal{A}_{\epsilon}$ from $\mathcal{A}$ by replacing
the unobservable label $\sigma \notin \Sigma_o$ with the empty label $\epsilon$ in each transition.
\STATE Construct the augmented TA $\overline{\mathcal{A}_{\epsilon}}$ by Definition \ref{def:ma}.
\STATE Build the region automaton
$\mathcal{R}(\overline{\mathcal{A}_{\epsilon}}) = (\Sigma_\epsilon, Q, Q_0, Q_f, \Gamma)$ by Definition \ref{def:ra}.

\STATE  Transform the NFA $\mathcal{R}(\overline{\mathcal{A}_{\epsilon}})$ to the DFA
$\mathscr{H} = ( \Sigma_{o}, \mathscr{X}, X_0, $ $ \mathscr{X}_f, \Upsilon )$ using the subset construction method.

\FOR{each state $X \in \mathscr{X}$ in $\mathscr{H}$}
  \STATE Let $L_X = \{ l \in L \mid \exists (l^+, R) \in X$ or $(l^0, R) \in X\}$.
  \IF{$L_X \cap L_s \neq \emptyset$ \textbf{and} $L_X \cap L_{ns} = \emptyset$}
      \RETURN ``NO". 
  \ENDIF
\ENDFOR
\RETURN ``YES". 
\end{algorithmic}
\end{algorithm}

\begin{theorem}
  Algorithm \ref{alg:clto} is correct.
\end{theorem}
\begin{IEEEproof}
  According to Definition \ref{def:clto}, $\mathcal{A}$ does not satisfy CLTO if and only if
  $P_{\Sigma_o}(\mathfrak{L}_{L_{0}}^{L_{s}}(\mathcal{A})) \not\subseteq
P_{\Sigma_o}(\mathfrak{L}_{L_{0}}^{L_{ns}}(\mathcal{A}))$, which is
further equivalent to $\mathfrak{L}_{L_{0}}^{L_{s}}(\mathcal{A_{\epsilon}}) \not\subseteq
 \mathfrak{L}_{L_{0}}^{L_{ns}}(\mathcal{A_{\epsilon}})$ because $\mathcal{A_{\epsilon}}$ is obtained from $\mathcal{A}$ by
 replacing the unobservable label with $\epsilon$ in each transition.
 Consider two $\epsilon$-IRTA $\mathcal{A_{\epsilon}^{'}}$ and $\mathcal{A_{\epsilon}^{''}}$
   obtained from $\mathcal{A_{\epsilon}}$ by replacing the accepting states with $L_s$ and $L_{ns}$, respectively.
 Then we have $\mathfrak{L}_{L_{0}}^{L_{s}}(\mathcal{A_{\epsilon}})  = \mathfrak{L}_{f}(\mathcal{A}_{\epsilon}^{'})$
 and  $\mathfrak{L}_{L_{0}}^{L_{ns}}(\mathcal{A_{\epsilon}}) = \mathfrak{L}_{f}(\mathcal{A}_{\epsilon}^{''})$.
In the region automaton $\mathcal{R}(\overline{\mathcal{A}_{\epsilon}})$ constructed in step 3 of Algorithm \ref{alg:clto},
define  $Q_s    = \{ (l^{*}, R) \in Q \mid  l \in  L_{s}, * \in \{0,+\} \}$
   and  $Q_{ns} = \{ (l^{*}, R) \in Q \mid  l \in L_{ns}, * \in \{0,+\} \}$.
Then by Definitions \ref{def:ra} and \ref{def:ma}, we obtain
  $\mathfrak{L}_{f}(\mathcal{R(\overline{A_{\epsilon}^{'}})}) = \mathfrak{L}_{Q_{0}}^{Q_{s}}(\mathcal{R(\overline{A_{\epsilon}})})$
  and
   $\mathfrak{L}_{f}(\mathcal{R(\overline{A_{\epsilon}^{''}})}) = \mathfrak{L}_{Q_{0}}^{Q_{ns}}(\mathcal{R(\overline{A_{\epsilon}})})$.
 By Lemma \ref{lemma:1}, we have
  $ \mathfrak{L}_f(\mathcal{A_{\epsilon}^{'}}) \not\subseteq \mathfrak{L}_f(\mathcal{A_{\epsilon}^{''}}) $ if and only if
  $ \mathfrak{L}_f(\mathcal{R(\overline{A_{\epsilon}^{'}})} ) \not\subseteq
  \mathfrak{L}_f(\mathcal{R(\overline{A_{\epsilon}^{''}})} )$.
  Thus, $\mathcal{A}$ does not satisfy CLTO
 if and only if
 $\mathfrak{L}_{Q_{0}}^{Q_{s}}(\mathcal{R(\overline{A_{\epsilon}})}) \not\subseteq
 \mathfrak{L}_{Q_{0}}^{Q_{ns}}(\mathcal{R(\overline{A_{\epsilon}})})$.

 \par
 $\mathfrak{L}_{Q_{0}}^{Q_{s}}(\mathcal{R(\overline{A_{\epsilon}})}) \not\subseteq
 \mathfrak{L}_{Q_{0}}^{Q_{ns}}(\mathcal{R(\overline{A_{\epsilon}})})$ holds if and only if
 there exists an observation $\omega \in (\Sigma_{o} \cup \checkmark \cup \delta)^{*}$
 that is generated by the run $q_0 \xrightarrow{} q_s$, where
 $q_{0} \in Q_{0}$ and $q_{s} \in Q_{s}$, and $\omega$ cannot be generated by a run
 $\overline{q_{0}} \xrightarrow{} q_{ns}$, where
 $\overline{q_{0}} \in Q_{0}$ and $q_{ns} \in Q_{ns}$, in $\mathcal{R(\overline{A_{\epsilon}})}$.
 Suppose that the observation $\omega$ is generated by the run $X_{0} \xrightarrow{} X$ in $\mathscr{H}$.
 Then we have $q_s \in X$ and $X \cap Q_{ns} = \emptyset$, which means that
 the location-projection of state $X$ (i.e., $L_X$) satisfies that
 $L_X \cap L_s \neq \emptyset$ and $L_X \cap L_{ns} = \emptyset$, and then
 Algorithm \ref{alg:clto} outputs ``NO".
 Therefore, $\mathcal{A}$ does not satisfy CLTO if and only if Algorithm \ref{alg:clto} outputs ``NO".
\end{IEEEproof}

\par

\begin{remark}
In Algorithm \ref{alg:clto},
the constructions of $\mathcal{A}_{\epsilon}$ and $\overline{\mathcal{A}_{\epsilon}}$
involve the straightforward manipulations of the transitions and locations, which requires $\mathcal{O}(2\cdot|L|+3\cdot|\Delta|)$ time.
 Because the number of  locations in $\overline{\mathcal{A}_{\epsilon}}$ is $2\cdot|L|$, and
 the fractional parts of all the clocks in an IRTA are always equal to each other (see Proposition 1 in \cite{IRTA}),
 the numbers of  regions and states in $\mathcal{R}(\overline{\mathcal{A}_{\epsilon}})$
 are at most $2 \cdot \prod_{c \in \mathcal{C}} (\kappa(c) + 1)$
 and  $4 \cdot |L| \cdot \prod_{c \in \mathcal{C}} (\kappa(c) + 1)$, respectively.
  Thus, constructing and searching the deterministic version of the NFA
  $\mathcal{R}(\overline{\mathcal{A}_{\epsilon}})$ (i.e., the DFA $\mathscr{H}$) requires
  $\mathcal{O}(|\Delta|\cdot 2^{4\cdot|L|\cdot\prod_{c\in\mathcal{C}}(\kappa(c)+1)})$ time.
  Hence, the overall computational complexity of Algorithm \ref{alg:clto} is
  $\mathcal{O}(|\Delta|\cdot 16^{|L|\cdot\prod_{c\in\mathcal{C}}(\kappa(c)+1)})$.
\end{remark}

\par
\begin{example}\label{ex:1}

Given an IRTA $\mathcal{A} = (\Sigma, L, L_0, L_f, \mathcal{C}, \Delta)$ with the set of observable events
$\Sigma_{o}=\{a\}$, the set of secret locations $L_s=\{l_{1}\}$, and the set of non-secret locations $L_{ns}=\{l_{3}\}$,
as shown in Fig. \ref{fig1}.
By Algorithm \ref{alg:clto}, we obtain $\overline{\mathcal{A}_{\epsilon}}$ and $\mathcal{R}(\overline{\mathcal{A}_{\epsilon}})$, as shown in Fig. \ref{fig2} and Fig. \ref{fig3}, respectively.
Finally, by the subset construction method, we obtain the DFA $\mathscr{H}$ of the NFA $\mathcal{R}(\overline{\mathcal{A}_{\epsilon}})$.
A part of $\mathscr{H}$ is shown in Fig. \ref{fig4}, in which the state $X=(l_1^{+},B)$ corresponds to the location-projection $L_{X} = \{l_1\}$ that satisfies $L_{X} \cap L_s \neq \emptyset$ and $L_X \cap L_{ns} = \emptyset$.
Hence, Algorithm \ref{alg:clto} outputs ``no", and $\mathcal{A}$ does not satisfy CLTO.
\par

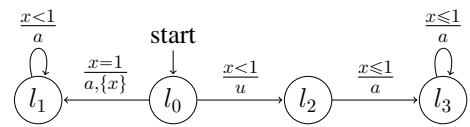
\begin{figure}[h]
\centering
 \resizebox{0.34\textwidth}{!}{%
          \begin{tikzpicture}[shorten >=1pt, node distance=2.5cm, auto, font=\Large,
          initial/.style={initial by arrow, initial where=above}]
          \node[state, initial] (l0) {$l_0$};
          \node[state, left of=l0] (l1) {$l_1$ }; 
          \node[state, right of=l0] (l2) {$l_2$};
          \node[state, right of=l2] (l3) {$l_3$ };

          \path[->, font=\Large]
            (l0) edge node[above] {$\frac{x=1}{a, \{x\}}$} (l1)  
            (l0) edge node[above]  {$\frac{x<1}{u }$} (l2)  
            (l2) edge node[above] {$\frac{x \leq 1}{a}$} (l3)
            (l3) edge[loop above] node {$\frac{x\leq 1}{a}$} ()
            (l1) edge[loop above] node {$\frac{x<1}{a}$} ();
        \end{tikzpicture}
    }
    \caption{The IRTA $\mathcal{A}$ in Example \ref{ex:1}.}
    \label{fig1}
\end{figure}

\begin{figure}[h]
\centering
 \resizebox{0.5\textwidth}{!}{%
        \begin{tikzpicture}[shorten >=1pt, node distance=3.10cm, auto,  font=\large,
        initial/.style={initial by arrow, initial where=above}]
          \node[state, initial] (l00) {$l_0^{0}$};
          \node[state, left of=l00] (l10) {$l_1^{0}$ }; 
          \node[state, right of=l00] (l20) {$l_2^{0}$};
          \node[state, right of=l20] (l30) {$l_3^{0}$ };

          \node[state, below of=l00, node distance=2.5cm] (l01) {$l_0^{+}$};
          \node[state, below of=l10, node distance=2.5cm] (l11) {$l_1^{+}$ }; 
          \node[state, below of=l20, node distance=2.5cm] (l21) {$l_2^{+}$};
          \node[state, below of=l30, node distance=2.5cm] (l31) {$l_3^{+}$ };

          \path[->,font=\Large]
            (l00) edge node[above] {$\frac{x=1 \wedge c=0}{a, \{x\}}$} (l10)
            (l00) edge node[above]  {$\frac{x<1 \wedge c=0}{\epsilon} $} (l20)
            (l20) edge node[above] {$\frac{x\leq1 \wedge c=0}{a}$} (l30)
            (l01) edge  node[below]  {$\frac{x<1\wedge 0<c<1}{\epsilon} $} (l21)
            (l21) edge node[below] {$\frac{x \leq 1 \wedge 0<c<1}{a}$} (l31)
            (l31) edge[loop below] node {$\frac{x\leq 1 \wedge 0<c<1}{a}$} ()
            (l11) edge[loop below] node {$\frac{x<1 \wedge 0<c<1}{a}$} ()
            (l30) edge[loop above] node {$\frac{x\leq 1 \wedge c=0}{a}$} ()
            (l10) edge[loop above] node {$\frac{x<1 \wedge c=0}{a}$} ()
            (l00) edge [bend left]node[ right]  {$\frac{0<c<1}{\delta} $} (l01)
            (l01) edge node[left] {$\frac{c=1}{\checkmark, \{c\}}$} (l00)
            (l10) edge[bend left] node[right] {$\frac{0<c<1}{\delta} $} (l11)
            (l11) edge node[left] {$\frac{c=1}{\checkmark, \{c\}}$} (l10)
            (l20) edge [bend left] node[ right] {$\frac{0<c<1 }{\delta}$} (l21)
            (l21) edge node[left] {$\frac{c=1}{\checkmark, \{c\}}$} (l20)
            (l30) edge [bend left]node[right] {$\frac{0<c<1 }{\delta}$} (l31)
            (l31) edge node[left] {$\frac{c=1}{\checkmark, \{c\}}$} (l30);
        \end{tikzpicture}
    }
    \caption{The $\overline{\mathcal{A}_{\epsilon}}$ in Example \ref{ex:1}, where the invalid transitions are erased.}
    \label{fig2}
\end{figure}
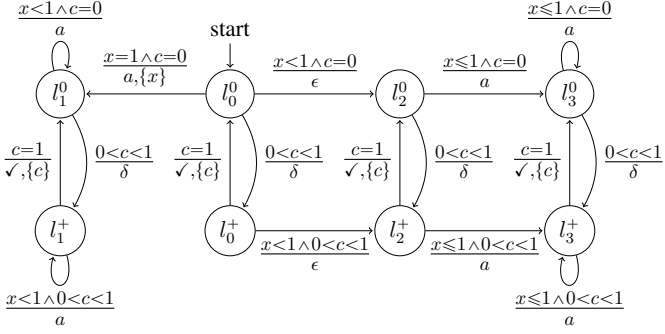

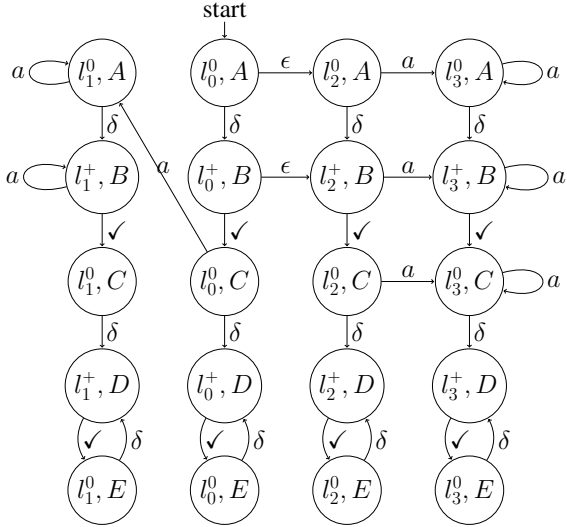
\begin{figure}[h]
    \centering
    \resizebox{0.42\textwidth}{!}{%
        \begin{tikzpicture}[shorten >=1pt, node distance=2.9cm, auto , font=\huge,
                initial/.style={initial by arrow, initial where=above}]
          \node[state, initial] (l00) {$l_0^{0},A$};
          \node[state, left of=l00, node distance=3.4cm] (l10) {$l_1^{0},A$ }; 
          \node[state, right of=l00, node distance=3.4cm] (l20) {$l_2^{0},A$};
          \node[state, right of=l20, node distance=3.4cm] (l30) {$l_3^{0},A$ };

          \node[state, below of=l00] (l01) {$l_0^{+},B$};
          \node[state, below of=l10] (l11) {$l_1^{+},B$ }; 
          \node[state, below of=l20] (l21) {$l_2^{+},B$};
          \node[state, below of=l30] (l31) {$l_3^{+},B$ };

          \node[state, below of=l01] (l02) {$l_0^{0},C$};
          \node[state, below of=l11] (l12) {$l_1^{0},C$ }; 
          \node[state, below of=l21] (l22) {$l_2^{0},C$};
          \node[state, below of=l31] (l32) {$l_3^{0},C$ };

          \node[state, below of=l02] (l03) {$l_0^{+},D$};
          \node[state, below of=l12] (l13) {$l_1^{+},D$ }; 
          \node[state, below of=l22] (l23) {$l_2^{+},D$};
          \node[state, below of=l32] (l33) {$l_3^{+},D$ };

          \node[state, below of=l03] (l04) {$l_0^{0},E$};
          \node[state, below of=l13] (l14) {$l_1^{0},E$ }; 
          \node[state, below of=l23] (l24) {$l_2^{0},E$};
          \node[state, below of=l33] (l34) {$l_3^{0},E$ };

          \path[->, font=\huge]
            (l00) edge node[above]  {$\epsilon $} (l20)
            (l20) edge node[above] {$a$} (l30)
            (l30) edge[loop right] node {$a$} ()
            (l10) edge[loop left] node {$a$} ()
            (l01) edge node[above]  {$\epsilon $} (l21)
            (l21) edge node[above] {$a$} (l31)
            (l31) edge[loop right] node {$a$} ()
            (l11) edge[loop left] node {$a$} ()
            (l22) edge node[above] {$a$} (l32)
            (l00) edge node[ right]  {$\delta $} (l01)
            (l10) edge node[right] {$\delta $} (l11)
            (l20) edge node[ right] {$\delta$} (l21)
            (l30) edge node[right] {$\delta $} (l31)
            (l01) edge node[right] {$\checkmark$} (l02)
            (l11) edge node[right] {$\checkmark$} (l12)
            (l21) edge node[right] {$\checkmark$} (l22)
            (l31) edge node[right] {$\checkmark$} (l32)
            (l02) edge node[above] {$a$} (l10)
            (l32) edge[loop right] node {$a$} ()
            (l02) edge node[ right]  {$\delta $} (l03)
            (l12) edge node[right] {$\delta $} (l13)
            (l22) edge node[ right] {$\delta$} (l23)
            (l32) edge node[right] {$\delta $} (l33)
            (l03) edge [bend right] node[right] {$\checkmark$} (l04)
            (l13) edge [bend right] node[right] {$\checkmark$} (l14)
            (l23) edge [bend right] node[right] {$\checkmark$} (l24)
            (l33) edge [bend right] node[right] {$\checkmark$} (l34)
            (l04) edge[bend right] node[right]  {$\delta $} (l03)
            (l14) edge[bend right] node[right] {$\delta $} (l13)
            (l24) edge[bend right] node[right] {$\delta$} (l23)
            (l34) edge[bend right] node[right] {$\delta $} (l33)
            ;
        \end{tikzpicture}
    }
    \caption{The $\mathcal{R}(\overline{\mathcal{A}_{\epsilon}})$ in Example \ref{ex:1}, where
   ``A",``B",``C",``D" and ``E" denote the regions of $[x=c=0]$, $[0<x=c<1]$, $[x=c=1]$, $[1 \leq x=c, \operatorname{frac}(c)=0]$
   and $[1\leq x=c, \operatorname{frac}(c) > 0]$, respectively, and the unreachable states are erased.}
    \label{fig3}
\end{figure}

\begin{figure}[!h]
    \centering
    \resizebox{0.48\textwidth}{!}{%
        \begin{tikzpicture}[font=\huge,->,>=stealth',shorten >=1pt,auto,semithick, node distance=3.0cm,
                                  every state/.style={rectangle,minimum width=0.23\textwidth, font=\huge},
                                  new-qs/.style={fill=gray!40,thick,font=\huge}]
                \node[initial, state] (x0) {$(l_0^{0},A), (l_2^{0},A)$};
                \node[state]  [ below of = x0] (x1) {$(l_0^{+},B), (l_2^{+},B)$};
                \node[state]  [node distance=6.0cm] [right of = x1] (x2) {$(l_0^{0},C), (l_2^{0},C)$};
                \node[state]  [node distance=6.0cm] [right of = x2] (x3) {$(l_1^{0},A), (l_3^{0},C) $};
                \node[state]  [above of = x3] (x4) {$(l_1^{+},B), (l_3^{+},D)$};
                \node[state,new-qs]  [node distance=6.0cm][left of = x4] (x5) {$(l_1^{+},B)$};

                \path[->, font=\huge]
                (x0) edge node[left]  {$\delta $} (x1)
                (x1) edge node[above]  {$\checkmark $} (x2)
                (x2) edge node[above]  {$a $} (x3)
                (x3) edge node[left]  {$\delta $} (x4)
                (x4) edge node[above]  {$a $} (x5)
                ;
        \end{tikzpicture}
    }
     \caption{A part of $\mathscr{H}$ in Example \ref{ex:1}.}
    \label{fig4}
\end{figure}
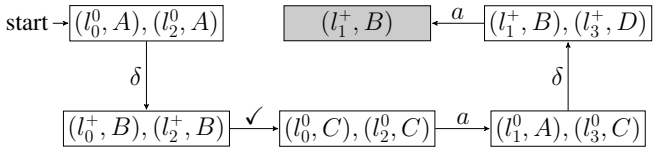

\end{example}

\par

\section{Timed Opacity against intruders with discrete-time  precision}

In this section, we establish decidable timed opacity by
weakening the observational capabilities of intruders.

\par
A substantial number of studies on timed opacity (e.g., \cite{wang2018-realtime-opacity}-\cite{Klein2024})
assume that intruders can perform exact real-valued clock measurements.
However, in some real-world applications, it may not be necessary to consider such a strong model of intruders,
as it is unlikely that an intruder possesses an exact real-valued clock in practice.
Hence, this section considers intruders with only discrete-time precision.

\par

Given the discrepancy in modeling granularity of clock valuations between systems and intruders,
it becomes a pivotal task in timed opacity analysis to estimate an intruder's observations for each timed word
$\omega = (\sigma_1, t_1)$ $(\sigma_2, t_2) \cdots (\sigma_n, t_n) \in \mathfrak{L}(\mathcal{A})$.

\par
First,
for a fractional timestamp $t_i$,
it is natural to assume that intruders shift $t_i$ to its preceding integer $\lfloor t_i \rfloor$ or
 following integer $\lceil t_i \rceil$ according to a specific threshold $\lambda \in [0,1)$,
 and the shifting operation should be order-preserving.
 For example, if $t_i$ ($t_i < 5$) is shifted to $5$, then any $  t_i' \in (t_i , 5]$ should also be shifted to $5$,
 and if $t_i$ ($t_i > 4$) is shifted to $4$, then any $  t_i' \in [4, t_i)$ should also be shifted to $4$.

  \par
  In practice, the threshold $\lambda$ is determined by the sensors used by intruders.
  Moreover, it is a challenge to estimate the thresholds of the potential sensors exactly,
  from the perspectives of both intruders and systems.
  Therefore, it is necessary to consider all possible $\lambda \in [0,1)$ in the observation estimation
  for intruders.

  \par
  Considering both the order-preserving property and the incalculability of the threshold as fundamental requirements,
  we define the timestamp \emph{shifting function} $S$ as follows:
  for  any timed word $ \omega = (\sigma_1, t_1)(\sigma_2, t_2) \cdots (\sigma_n, t_n) \in \mathfrak{L}(\mathcal{A})$,
  \begin{align}
    & S(\omega) = \{(\sigma_1, t_1^{\lambda})(\sigma_2, t_2^{\lambda}) \cdots (\sigma_n, t_n^{\lambda})
    \mid \lambda \in [0,1) \}, \nonumber \\
    & \text{ where } t_i^{\lambda} = \begin{cases}
                     \lfloor t_i \rfloor & \text{if } \operatorname{frac}(t_i) \leq \lambda \\
                     \lceil t_i \rceil   & \text{otherwise}
                    \end{cases} , i \in [1:n].
  \end{align}
  $S$ can be extended to timed languages in the usual way.
  This function is also referred to as \emph{digitization} in \cite{digitization1}-\cite{digitization3}.

   \par
Second, intruders cannot observe the unobservable timed events $(\sigma_i, t_i)$ with $\sigma_i \not\in \Sigma_{o}$.
This can be characterized by the projection function $P_{\Sigma_o}$ (see Eq. \ref{eq:observation}), similar to the conventional situation.

\par
Therefore, for any timed word $\omega \in \mathfrak{L}(\mathcal{A})$,
an intruder's observation $\text{Obs}(\omega)$ can be computed by the composition of
the projection function and the shifting function: $\text{Obs}(\omega) = P_{\Sigma_{o}}( S (\omega))$.
Note that the operations of these two functions are commutative, that is,
$P_{\Sigma_{o}}( S(\omega)) = S(P_{\Sigma_{o}}(\omega))$.

\par

Based on the aforementioned discussion,
we can define the notion of
\emph{Current-Location Timed Opacity against Intruders with Discrete-Time Precision} (CLTO-IDTP) as follows.

\begin{definition} \label{def:clto2}
Given a TA $\mathcal{A}$, an observable event set $\Sigma_o \subseteq \Sigma$,
a set of secret locations $L_s \subseteq L$,
and a set of non-secret locations $L_{ns} \subseteq L$,
$\mathcal{A}$ is \emph{current-location timed opaque against intruders with discrete-time precision} with respect to $\Sigma_o$, $L_s$, and $L_{ns}$, if $ \forall \omega \in \mathfrak{L}_{L_{0}}^{L_{s}}(\mathcal{A})$,
\begin{align}\label{eq:clto}
 \exists \omega' \in \mathfrak{L}_{L_{0}}^{L_{ns}}(\mathcal{A}) \text{ such that }
          S(P_{\Sigma_o}(\omega)) = S(P_{\Sigma_o}(\omega')),
\end{align}
or equivalently $S(P_{\Sigma_o}(\mathfrak{L}_{L_{0}}^{L_{s}}(\mathcal{A}))) \subseteq
S(P_{\Sigma_o}(\mathfrak{L}_{L_{0}}^{L_{ns}}(\mathcal{A})))$.
\end{definition}
\par
\begin{corollary} \label{corollary1}
  Given a TA $\mathcal{A}$, if $\mathcal{A}$ satisfies CLTO, then it also satisfies CLTO-IDTP.
\end{corollary}
\begin{IEEEproof}
   Follows straightforwardly from the definitions of the shifting function $S$, CLTO, and CLTO-IDTP.
\end{IEEEproof}

\par
Corollary \ref{corollary1} captures our intuition: a system resisting strong intruders inherently resists weaker ones.
\par

\begin{corollary} \label{corollary2}
  Given a TA $\mathcal{A}$, even if $\mathcal{A}$ satisfies CLTO-IDTP,
  $\mathcal{A}$ does \emph{not} necessarily satisfy CLTO.
\end{corollary}
\begin{IEEEproof}
  By contradiction, we assume that for any TA $\mathcal{A}$,
  if $\mathcal{A}$ satisfies CLTO-IDTP, then $\mathcal{A}$ also satisfies CLTO.
 That is,  CLTO-IDTP $\Rightarrow$ CLTO holds universally across all TA.
 In Corollary \ref{corollary1}, we showed that CLTO $\Rightarrow$ CLTO-IDTP.
 Therefore, the properties CLTO-IDTP and CLTO are equivalent for a general TA.
 On the other hand, CLTO has been proved to be undecidable in \cite{timed-opacity},
 yet CLTO-IDTP is decidable (we will present a verification algorithm in the rest of this section).
 This result contradicts the equivalence between CLTO-IDTP and CLTO.
\end{IEEEproof}

\par
In the following, we investigate the verification of CLTO-IDTP.
Firstly, we present a finite
structure to characterize the shifting operation on the language of a TA.
\par
\begin{definition} \label{def:ctr}
  Given a TA $\mathcal{A}= ( \Sigma, L, L_0, L_f, \mathcal{C}, \Delta )$,
  the \emph{closed timed region automaton} is a TA
$\mathcal{CTR}({\mathcal{A}}) =(\Sigma, Q,Q_0, $ $ Q_f, \Delta')$,
where $\Sigma, Q, Q_0, Q_f$ are the same as those in region automaton,
and the transition relation
$\Delta' \subseteq Q \times \Sigma \times \Phi(\mathcal{C}) \times 2^{\mathcal{C}} \times Q$ is defined as:
$((l, R), \sigma, \phi', r,  (l', R')) \in \Delta'$
if and only if there exists a region $R^{''}$ and
$(l, \sigma, \phi, r, l') \in \Delta$ such that
\begin{itemize}
  \item $R \preceq R^{''}$ and $R^{''}\models \phi$,
  \item $R' = \{ [r \mapsto 0]v \mid v \in R^{''} \}$,
  \item $\phi'$ is obtained from $\phi$ by
   replacing every strict inequality (e.g., ``$c<k$", $c \in \mathcal{C}$, $k \in \mathbb{N}$) by the corresponding non-strict one (e.g., ``$c \leq k$").
\end{itemize}
\end{definition}

\par
By means of the integral automata (Definition \ref{def:integralautomaton}) and
the closed timed region automata (Definition \ref{def:ctr}),
we present the verification algorithm for CLTO-IDTP (Algorithm \ref{alg:clto-idtp}).

\par

\begin{algorithm}[h]
\caption{Verification of Current-Location Timed Opacity against Intruders with Discrete-Time Precision (CLTO-IDTP)}
\label{alg:clto-idtp}
\textbf{Input}: A TA $\mathcal{A} = (\Sigma, L, L_0, L_f, \mathcal{C}, \Delta)$,
a set of observable events $\Sigma_o \subseteq \Sigma$,
a set of secret locations $L_s \subseteq L$, and a set of non-secret locations $L_{ns} \subseteq L$.

\textbf{Output}: ``YES" if $\mathcal{A}$ satisfies CLTO-IDTP; 
                 ``NO" otherwise.

\begin{algorithmic}[1]
\STATE Obtain $\mathcal{A}_{\epsilon}$ from $\mathcal{A}$ by replacing $\sigma \notin \Sigma_o$
with $\epsilon$ in each transition.
\STATE  Compute the closed timed region automaton
$\mathcal{CTR}({\mathcal{A}_{\epsilon}})  =(\Sigma_{\epsilon}, Q,Q_0, Q_f, \Delta')$
according to Definition \ref{def:ctr}.
\STATE Compute the reduction TA $\mathcal{B} = (\Sigma^{B}, Q^{B},Q_0^{B}, Q_f^{B}, \Delta^{B})$
for $\mathcal{CTR}({\mathcal{A}_{\epsilon}})$ by Algorithm \ref{alg:reduction}.

\STATE  Compute the integral automaton
$\mathcal{B}^{\checkmark} = (\Sigma_{\epsilon} \cup \{\checkmark\}, Q^\checkmark, $ $ Q_0^\checkmark, Q_f^\checkmark,
\Delta^\checkmark)$
for $\mathcal{B}$ according to Definition \ref{def:integralautomaton}.

\STATE  Transform the NFA $\mathcal{B}^{\checkmark}$ to the DFA
$\mathscr{H} = ( \Sigma_{o}\cup \{\checkmark\}, \mathscr{X}, $ $ X_0, \mathscr{X}_f, \Upsilon )$
using the subset construction method.

\FOR{each state $X \in \mathscr{X}$ in $\mathscr{H}$}
  \STATE Let $\overline{L}_X = \{ l \in L \mid \exists ((l, R_1),R_2) \in X \}$
  \IF{$\overline{L}_X \cap L_s \neq \emptyset$ \textbf{and} $\overline{L}_X \cap L_{ns} = \emptyset$}
      \RETURN ``NO" 
  \ENDIF
\ENDFOR
\RETURN ``YES". 
\end{algorithmic}
\end{algorithm}

\par

\begin{algorithm} \label{alg:reduction}
\caption{Reduce TA through Forward and Backward Simulation Relations Computation}
\textbf{Input}: A TA $\mathcal{A} = (\Sigma, L, L_0, L_f, \mathcal{C}, \Delta)$ and
its closed timed region automaton  $\mathcal{CTR}({\mathcal{A}_{\epsilon}}) =(\Sigma, Q,Q_0, Q_f, \Delta')$.
\par
\textbf{Output}: The TA $\mathcal{B} =(\Sigma^{B}, Q^{B},Q_0^{B}, Q_f^{B}, \Delta^{B})$
equivalent to $\mathcal{CTR}({\mathcal{A}_{\epsilon}})$. 
\par
\begin{algorithmic}[1]\label{alg:reduction}
    \STATE \textbf{Initialize} $\mathcal{B} \gets \mathcal{CTR}({\mathcal{A}_{\epsilon}})$.
        \STATE \textbf{Define} the \emph{forward simulation relation} $\preceq_l^{f}$ over $Q^{B}$ w.r.t.  $l$:
          $\{ ((l,R_2),(l, R_1)) \mid \forall ((l, R_2), \sigma, {\phi},  r, (l', R'_{2})) \in \Delta'$,
          $ \exists  ((l, R_1), \sigma, {\phi}, r,  (l', R'_{1})) \in \Delta'
          \text{ and } (l', R'_{2}) \preceq_{l'}^{f} (l', R'_{1}) \}$.
          \textbf{Compute} the maximal $\preceq_l^{f}$ for each $l \in L$.
        \STATE \textbf{Define} the \emph{backward simulation relation} $\preceq_l^{b}$ over $Q^{B}$ w.r.t.  $l$:
           $\{ ((l,R_2),(l, R_1)) \mid \forall ((l', R'_{2}), \sigma, {\phi},  r,(l, R_2)) \in \Delta'$,
           $ \exists  ((l', R'_{1}), \sigma, {\phi}, r, (l, R_1)) \in \Delta'
           \text{ and } (l', R'_{2}) \preceq_{l'}^{b} (l', R'_{1}) \}$.
           \textbf{Compute} the maximal $\preceq_l^{b}$ for each $l \in L$.
    \FOR{each location $l \in L$}
        \STATE For each $(l, R_2) \in Q^{B}\backslash Q_0^{B}$, if there exists $(l, R_1) \in Q$ such that
        $(l, R_2) \preceq_l^{f} (l, R_1)$ and $(l, R_2) \preceq_l^{b} (l, R_1)$, then
        remove $(l, R_2)$ and its corresponding transitions from $\mathcal{B}$.
    \ENDFOR
    \STATE \textbf{Return} the remaining part of $\mathcal{B}$.
\end{algorithmic}
\end{algorithm}

\par
Before providing a formal proof for the correctness of Algorithm \ref{alg:clto-idtp},
we introduce three necessary lemmas in the following.
\par
First, it has been proved in \cite{digitization2} that the closed timed region automaton $\mathcal{CTR}({\mathcal{A}})$
exactly characterizes, by its integral accepting  language, the shifting operation on
the accepting language of the TA $\mathcal{A}$.
\par
\begin{lemma} \label{lemma:ctr}
  For a TA $\mathcal{A}$, 
  $S(\mathfrak{L}_{f}(\mathcal{A})) =
  (\mathfrak{L}_{f}(\mathcal{CTR}({\mathcal{A}})) \cap (\Sigma \times \mathbb{N})^{*})$.
\end{lemma}

\par
Second, according to Definition \ref{def:integralautomaton},
an integral timed word in the  language of a TA $\mathcal{A}$ is in
one-to-one correspondence with an untimed word in the  language of its integral automaton
$\mathcal{A}^{\checkmark}$.
Hence, we have the following lemma immediately.
\begin{lemma}\label{lemma:tick}
  Given two TA $\mathcal{A}_{1}$ and $\mathcal{A}_{2}$,
  $(\mathfrak{L}_{f}(\mathcal{A}_{1}) \cap (\Sigma \times \mathbb{N})^{*}) \subseteq
   (\mathfrak{L}_{f}(\mathcal{A}_{2}) \cap (\Sigma \times \mathbb{N})^{*}) \Longleftrightarrow
    \mathfrak{L}_{f}(\mathcal{A}_{1}^{\checkmark}) \subseteq \mathfrak{L}_{f}(\mathcal{A}_{2}^{\checkmark})$.
\end{lemma}

\par
Third, we prove that the TA $\mathcal{B}$ computed by Algorithm \ref{alg:reduction} is exactly the reduction of the closed
timed region automaton $\mathcal{CTR}({\mathcal{A}_{\epsilon}})$.
\par
\begin{lemma}\label{lemma:reduction}
Given the $\mathcal{CTR}({\mathcal{A}_{\epsilon}})$ and its reduction TA $\mathcal{B}$,
let $Q_{s} = \{(l,R) \in Q \mid l \in L_{s}\}$ and $Q_{ns} = \{(l,R) \in Q \mid l \in L_{ns}\}$.
  $\mathfrak{L}_{f}(\mathcal{B}) = \mathfrak{L}_{f}(\mathcal{CTR}({\mathcal{A}_{\epsilon}}))$,
  $\mathfrak{L}_{Q_{0}^{B}}^{Q_{s}}(\mathcal{B}) = \mathfrak{L}_{Q_{0}}^{Q_{s}}(\mathcal{CTR}({\mathcal{A}_{\epsilon}}))$,  and
  $\mathfrak{L}_{Q_{0}^{B}}^{Q_{ns}}(\mathcal{B}) = \mathfrak{L}_{Q_{0}}^{Q_{ns}}(\mathcal{CTR}({\mathcal{A}_{\epsilon}}))$.
\end{lemma}
\begin{IEEEproof}
  According to Algorithm \ref{alg:reduction}, $\mathcal{B}$ is obtained from $\mathcal{CTR}({\mathcal{A}_{\epsilon}})$
  by removing every non-initial state $(l,R_2)$ that can be simulated by $(l,R_1)$ both forward and backward.
  Firstly, $(l, R_2) \preceq_l (l, R_1)$ and $(l, R_2) \preceq_l^{b} (l, R_1)$
  mean that $(l, R_1)$ can simulate all past and future behaviors of $(l, R_2)$, that is,
   any timed word visiting $(l, R_2)$ also visits $(l, R_1)$.
  Moreover, the removed state $(l, R_2)$ is a non-initial state,
  and $(l, R_2)$ and $(l, R_1)$ share the same accepting and secrecy attributes.
Therefore, the operation of removing state $(l, R_2)$ does not alter the accepting, secret, and non-secret languages of
$\mathcal{CTR}({\mathcal{A}_{\epsilon}})$.
\end{IEEEproof}

\begin{theorem}
  Algorithm \ref{alg:clto-idtp} is correct.
\end{theorem}
\begin{IEEEproof}
 By Definition \ref{def:clto2}, $\mathcal{A}$ does not satisfy CLTO-IDTP
  if and only if
  $S(P_{\Sigma_o}(\mathfrak{L}_{L_{0}}^{L_{s}}(\mathcal{A}))) \not\subseteq
S(P_{\Sigma_o}(\mathfrak{L}_{L_{0}}^{L_{ns}}(\mathcal{A})))$, which means that
 $S(\mathfrak{L}_{L_{0}}^{L_{s}}(\mathcal{A_{\epsilon}})) \not\subseteq
 S(\mathfrak{L}_{L_{0}}^{L_{ns}}(\mathcal{A_{\epsilon}}))$ because
 $\mathcal{A_{\epsilon}}$ exactly characterizes the projection operation on $\mathcal{A}$.
 Let $\mathcal{A_{\epsilon}^{'}}$ and $\mathcal{A_{\epsilon}^{''}}$ be two $\epsilon$-TA derived from
 $\mathcal{A_{\epsilon}}$ by replacing the set of accepting states with $L_s$ and $L_{ns}$, respectively.
 Then we have $\mathfrak{L}_{L_{0}}^{L_{s}}(\mathcal{A_{\epsilon}})  = \mathfrak{L}_{f}(\mathcal{A}_{\epsilon}^{'})$
 and  $\mathfrak{L}_{L_{0}}^{L_{ns}}(\mathcal{A_{\epsilon}}) = \mathfrak{L}_{f}(\mathcal{A}_{\epsilon}^{''})$.
 According to Lemma \ref{lemma:ctr}, we have
 $S(\mathfrak{L}_{f}(\mathcal{A}_{\epsilon}^{'})) =
  (\mathfrak{L}_{f}(\mathcal{CTR}({\mathcal{A}_{\epsilon}^{'})})) \cap (\Sigma \times \mathbb{N})^{*})$ and
   $S(\mathfrak{L}_{f}(\mathcal{A}_{\epsilon}^{''})) =
  (\mathfrak{L}_{f}(\mathcal{CTR}({\mathcal{A}_{\epsilon}^{''})})) \cap (\Sigma \times \mathbb{N})^{*})$.
  Therefore, $\mathcal{A}$ does \emph{not} satisfy CLTO-IDTP if and only if
  $ (\mathfrak{L}_{f}(\mathcal{CTR}({\mathcal{A}_{\epsilon}^{'})})) \cap (\Sigma \times \mathbb{N})^{*}) \not\subseteq
    (\mathfrak{L}_{f}(\mathcal{CTR}({\mathcal{A}_{\epsilon}^{''})})) \cap (\Sigma \times \mathbb{N})^{*})$.

  \par
  In the reduced TA $\mathcal{B}$
  computed by Algorithm \ref{alg:reduction},
 we define  $Q_s^{B}    = \{ (l, R_1) \in Q^{B} \mid  l \in  L_{s}\}$
  and  $Q_{ns}^{B} = \{ (l, R_1) \in Q^{B} \mid  l \in L_{ns}\}$,
  and further define $\mathcal{B}_{1}$ and $\mathcal{B}_{2}$ from $\mathcal{B}$ by replacing the set of accepting states
  with $Q_s^{B}$ and $Q_{ns}^{B}$, respectively.
  According to Definition \ref{def:ctr} and Lemma \ref{lemma:reduction}, we have
  $\mathfrak{L}_{f}(\mathcal{B}_{1}) = \mathfrak{L}_{f}(\mathcal{CTR}({\mathcal{A}^{'}_{\epsilon}}))$ and
  $\mathfrak{L}_{f}(\mathcal{B}_{2}) = \mathfrak{L}_{f}(\mathcal{CTR}({\mathcal{A}^{''}_{\epsilon}}))$.
  By Lemma \ref{lemma:tick},
  $(\mathfrak{L}_{f}(\mathcal{B}_{1}) \cap (\Sigma \times \mathbb{N})^{*}) \not\subseteq
  (\mathfrak{L}_{f}(\mathcal{B}_{2}) \cap (\Sigma \times \mathbb{N})^{*})$ is equivalent to
  $\mathfrak{L}_{f}(\mathcal{B}_{1}^{\checkmark}) \not\subseteq \mathfrak{L}_{f}(\mathcal{B}_{2}^{\checkmark})$.
  Define $Q_{s}^{\checkmark} = \{ (q_{s}^{B},R) \in Q^{\checkmark} \mid q_{s}^{B} \in Q_{s}^{B} \}$ and
  $Q_{ns}^{\checkmark} = \{ (q_{ns}^{B},R) \in Q^{\checkmark} \mid q_{ns}^{B} \in Q_{ns}^{B} \}$,
  then by Definition \ref{def:integralautomaton}, we have
  $\mathfrak{L}_{f}(\mathcal{B}_{1}^{\checkmark}) =
   \mathfrak{L}_{Q_{0}^{\checkmark}}^{Q_{s}^{\checkmark}}(\mathcal{B}^{\checkmark})$
  and
    $\mathfrak{L}_{f}(\mathcal{B}_{2}^{\checkmark}) =
   \mathfrak{L}_{Q_{0}^{\checkmark}}^{Q_{ns}^{\checkmark}}(\mathcal{B}^{\checkmark})$.
 Therefore, $\mathcal{A}$ does not satisfy CLTO-IDTP if and only if
  $\mathfrak{L}_{Q_{0}^{\checkmark}}^{Q_{s}^{\checkmark}}(\mathcal{B}^{\checkmark}) \not\subseteq
  \mathfrak{L}_{Q_{0}^{\checkmark}}^{Q_{ns}^{\checkmark}}(\mathcal{B}^{\checkmark})$.

 \par
 $\mathfrak{L}_{Q_{0}^{\checkmark}}^{Q_{s}^{\checkmark}}(\mathcal{B}^{\checkmark}) \not\subseteq
  \mathfrak{L}_{Q_{0}^{\checkmark}}^{Q_{ns}^{\checkmark}}(\mathcal{B}^{\checkmark})$ holds if and only if
 there exists an observation
 $\omega \in (\Sigma_{o}\cup \{\checkmark\})^{*}$ that can be generated by the run
 $q_0^{\checkmark} \xrightarrow{} q_s^{\checkmark}$, where
 $q_{0}^{\checkmark} \in Q_{0}^{\checkmark}$ and $q_{s}^{\checkmark} \in Q_{s}^{\checkmark}$,
 and $\omega$ cannot be generated by any run ${p_{0}^{\checkmark}} \xrightarrow{} p_{ns}^{\checkmark}$, where
 ${p_{0}^{\checkmark}} \in Q_{0}^{\checkmark}$ and $p_{ns}^{\checkmark} \in Q_{ns}^{\checkmark}$,
 in $\mathcal{B}^{\checkmark}$.
 Suppose the observation $\omega$ is generated by the run $X_{0} \xrightarrow{} X$ in $\mathscr{H}$.
 Then we have $q_s^{\checkmark} \in X$ and $X \cap Q_{ns}^{\checkmark} = \emptyset$,
   which means that
 $\overline{L}_X \cap L_s \neq \emptyset$ and $\overline{L}_X \cap L_{ns} = \emptyset$,
 and Algorithm \ref{alg:clto-idtp} outputs ``NO".
 Hence, $\mathcal{A}$ does not satisfy CLTO-IDTP if and only if Algorithm \ref{alg:clto-idtp} outputs ``NO".
\end{IEEEproof}

\begin{remark}
For the analysis of the computational complexity of Algorithm \ref{alg:clto-idtp},
the critical steps being considered include the following:
 the construction and reduction of the closed timed region automaton $\mathcal{CTR}(\mathcal{A}_{\epsilon})$,
  the determinization of $\mathcal{B}^{\checkmark}$, and the search of the DFA $\mathscr{H}$.
 First, constructing $\mathcal{CTR}(\mathcal{A}_{\epsilon})$ involves generating at most $|L|\cdot|\mathcal{C}|!\cdot4^{|\mathcal{C}|}\cdot\prod_{c\in\mathcal{C}}(\kappa(c)+1)$ states and
  $|\Delta|\cdot|L|\cdot|\mathcal{C}|!\cdot4^{|\mathcal{C}|}\cdot\prod_{c\in\mathcal{C}}(\kappa(c)+1)$ transitions,
  which requires $O(|\Delta|\cdot|L|\cdot|\mathcal{C}|!\cdot4^{|\mathcal{C}|}\cdot\prod_{c\in\mathcal{C}}(\kappa(c)+1))$
  time in the worst case.
  Since $\mathcal{CTR}(\mathcal{A}_{\epsilon})$ is excessively large with numerous redundant states
  for subsequent processes,
  it is reduced to $\mathcal{B}$
  through the maximum forward and backward simulation relations computation.
  The state-of-the-art algorithm for maximum simulation computation (see \cite{simulation})
  requires
  $\mathcal{O}(|\Delta|\cdot|L|\cdot|\mathcal{C}|!\cdot4^{|\mathcal{C}|}\cdot\prod_{c\in\mathcal{C}}(\kappa(c)+1))$ time.
  Suppose the number of states in $\mathcal{B}$ is $|Q^{B}|$.
  Constructing $\mathcal{B}^{\checkmark}$ generates at most $|Q^{B}|\cdot\prod_{c\in\mathcal{C}}(\kappa(c)+1)$ states
  and $|\Delta|\cdot|Q^{B}|\cdot\prod_{c\in\mathcal{C}}(\kappa(c)+1)$ transitions,
  which takes $\mathcal{O}(|\Delta|\cdot|Q^{B}|\cdot\prod_{c\in\mathcal{C}}(\kappa(c)+1))$ time.
  Constructing and searching $\mathscr{H}$ consume
  $\mathcal{O}(|\Delta|\cdot 2^{|Q^{B}|\cdot\prod_{c\in\mathcal{C}}(\kappa(c)+1)})$ time in the worst case.
  Therefore, the overall computational complexity of Algorithm \ref{alg:clto-idtp} is
  $\mathcal{O}(|\Delta|\cdot(|L|\cdot|\mathcal{C}|!\cdot4^{|\mathcal{C}|}\cdot\prod_{c\in\mathcal{C}}(\kappa(c)+1) + 2^{|Q^{B}|\cdot\prod_{c\in\mathcal{C}}(\kappa(c)+1)}))$.
\end{remark}

\begin{example} \label{ex:2}
 Given a TA $\mathcal{A} = (\Sigma, L, L_0, L_f, \mathcal{C}, \Delta)$ with the set of observable events
  $\Sigma_{o}=\{a, b\}$, the set of secret locations $L_s=\{l_{3}\}$ and the set of non-secret locations $L_{ns}=\{l_{4}\}$,
   as shown in Fig. \ref{fig5}.
   By  steps 2 and 3 of Algorithm \ref{alg:clto-idtp},
   we obtain the closed region automaton $\mathcal{CTR}({\mathcal{A}_{\epsilon}})$ and
   its reduction TA $\mathcal{B}$,
   as shown in Fig. \ref{fig6} and Fig. \ref{fig7}, respectively.
   Following  step 4 of Algorithm \ref{alg:clto-idtp},
   we construct the integral automaton $\mathcal{B}^{\checkmark}$, as shown in Fig. \ref{fig8}.
Finally, by the subset construction method, we obtain the DFA $\mathscr{H}$ of the NFA $\mathcal{B}^{\checkmark}$.
A part of $\mathscr{H}$ is shown in Fig. \ref{fig9}, in which only the reachable states $X$ satisfying
$\overline{L}_{X} \cap {L}_{s} \neq \emptyset$ or the states that can reach those states $X$ are shown.
     We  find that each state $X$ satisfying $\overline{L}_{X} \cap L_{s} \neq \emptyset$ also satisfies
     $\overline{L}_{X} \cap L_{ns} \neq \emptyset$.
     Hence, Algorithm \ref{alg:clto-idtp} outputs ``yes", and $\mathcal{A}$  satisfies CLTO-IDTP.

\par
    However, it is easy to verify that $\mathcal{A}$ does not satisfy CLTO.
    This is because every timed word reachable to the secret location $l_3$
    produces observations starting with the timed event $(a,1)$.
    However, every timed word reachable to the non-secret location $l_4$ cannot produce such observations.
    Hence, this example also serves as an illustration to demonstrate the correctness of Corollary \ref{corollary2}.

\begin{figure}[!h]
\centering
 \resizebox{0.41\textwidth}{!}{%
          \begin{tikzpicture}[shorten >=1pt, node distance=2.5cm, auto, font=\large,
          initial/.style={initial by arrow, initial where=above},
          new-qs/.style={fill=gray!40,thick}]
          \node[state, initial] (l0) {$l_0$};
          \node[state, left of=l0] (l1) {$l_1$ };
          \node[state, left of=l1] (l4) {$l_4$ }; 
          \node[state, right of=l0] (l2) {$l_2$};
          \node[state,  right of=l2] (l3) {$l_3$ };

          \path[->,font=\LARGE]
            (l0) edge node[above] {$\frac{x > 1}{a, \{x\} }$} (l1)  
            (l0) edge node[above]  {$\frac{x = 1}{a }$} (l2)  
            (l2) edge  node[below] {$\frac{x < 1}{a}$} (l3)
            (l2) edge [bend left]  node[above] {$\frac{x > 1}{b, \{x\}}$} (l3)
            (l1) edge node[above] {$\frac{x \leq 1}{u}$} (l4)
            (l3) edge[loop above] node {$\frac{x > 1}{b,\{x\}}$} ()
            (l4) edge[loop above] node {$\frac{x > 0}{b,\{x\}}$} ();
        \end{tikzpicture}
    }
    \caption{The TA $\mathcal{A}$ in Example \ref{ex:2}.}
    \label{fig5}
\end{figure}
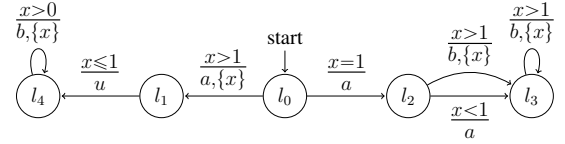

\begin{figure}[!h]
\centering
 \resizebox{0.48\textwidth}{!}{%
          \begin{tikzpicture}[shorten >=1pt, node distance=2.9cm, auto, font=\Large,
          initial/.style={initial by arrow, initial where=above}]
          \node[state, initial] (l0) {$(l_0,\widetilde{A})$};
          \node[state, left of=l0] (l1) {$(l_1,\widetilde{A})$ };
          \node[state, left of=l1] (l4) {$(l_4,\widetilde{A})$ }; 
          \node[state, right of=l0] (l2) {$(l_2,\widetilde{C})$};
          \node[state, right of=l2] (l3) {$(l_3,\widetilde{A})$ };
          \node[state, below of=l4] (l4b) {$(l_4,\widetilde{B})$ }; 
          \node[state, above of=l4] (l4c) {$(l_4,\widetilde{C})$ }; 

          \path[->, font=\huge]
            (l0) edge node[above] {$\frac{x \geq 1}{a, \{x\} }$} (l1)  
            (l0) edge node[above]  {$\frac{x = 1}{a }$} (l2)  
            (l2) edge  node[above] {$\frac{x \geq 1}{b, \{x\}}$} (l3)
            (l1) edge node[right, pos=0.8, align=right, xshift=3mm] {$\frac{x \leq 1}{\epsilon}$} (l4b)
            (l1) edge node[right] {$\frac{x \leq 1}{\epsilon}$} (l4c)
            (l1) edge node[above] {$\frac{x \leq 1}{\epsilon}$} (l4)
            (l4b) edge node {$\frac{x \geq 0}{b,\{x\}}$} (l4)
            (l4c) edge node[left] {$\frac{x \geq 0}{b,\{x\}}$} (l4)
            (l3) edge[loop above] node {$\frac{x \geq 1}{b,\{x\}}$} ()
            (l4) edge[loop left] node {$\frac{x \geq 0}{b,\{x\}}$} ();
        \end{tikzpicture}
    }
    \caption{The closed region automaton $\mathcal{CTR}({\mathcal{A}_{\epsilon}})$ in Example \ref{ex:2},  where
   ``$\widetilde{A}$",``$\widetilde{B}$", and ``$\widetilde{C}$" denote the regions of $[x=0]$, $[0<x<1]$, $[x=1]$, respectively, and all unreachable states are erased.}
    \label{fig6}
\end{figure}
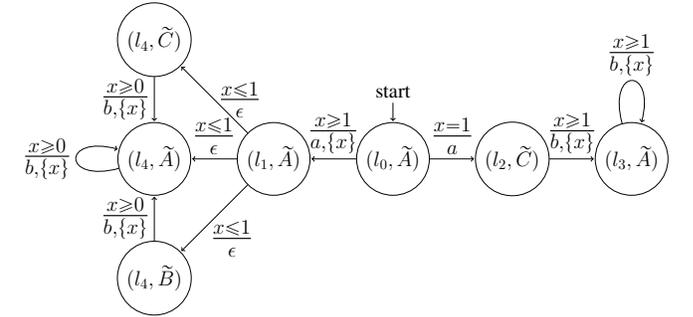

\begin{figure}[!h]
\centering
 \resizebox{0.4\textwidth}{!}{%
          \begin{tikzpicture}[shorten >=1pt, node distance=2.5cm, auto, font=\large,
          initial/.style={initial by arrow, initial where=above}]
          \node[state, initial] (l0) {$Q_0$};
          \node[state, left of=l0] (l1) {$Q_1$ }; 
          \node[state, left of=l1] (l4) {$Q_4$ }; 
          \node[state, right of=l0] (l2) {$Q_2$}; 
          \node[state, right of=l2] (l3) {$Q_3$ };

          \path[->, font=\LARGE]
            (l0) edge node[above] {$\frac{x \geq 1}{a, \{x\} }$} (l1)  
            (l0) edge node[above]  {$\frac{x = 1}{a }$} (l2)  
            (l2) edge  node[above] {$\frac{x \geq 1}{b, \{x\}}$} (l3)
            (l1) edge node[above] {$\frac{x \leq 1}{\epsilon}$} (l4)
            (l3) edge[loop above] node {$\frac{x \geq 1}{b, \{x\}}$} ()
            (l4) edge[loop above] node {$\frac{x \geq 0}{b, \{x\}}$} ();
        \end{tikzpicture}
    }
    \caption{The reduction automaton $\mathcal{B}$ in Example \ref{ex:2}, where $Q_4$, $Q_1$, $Q_0$, $Q_2$ and $Q_3$ denote the states $(l_4, \widetilde{A})$, $(l_1, \widetilde{A})$,$(l_0, \widetilde{A})$,$(l_2, \widetilde{C})$,
     and $(l_3, \widetilde{A})$, respectively, in Fig. \ref{fig6}.}
    \label{fig7}
\end{figure}
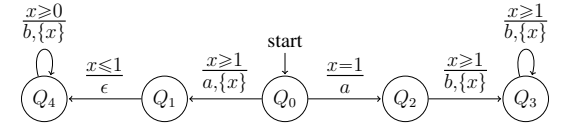

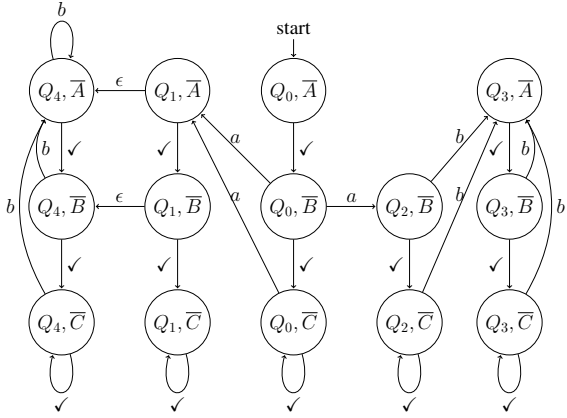
\begin{figure}[!h]
\centering
 \resizebox{0.42\textwidth}{!}{%
          \begin{tikzpicture}[shorten >=1pt, node distance=2.9cm, auto, font=\Large,
          initial/.style={initial by arrow, initial where=above}]
          \node[state, initial] (l0) {$Q_0,\overline{A}$};
          \node[state, left of=l0] (l1) {$Q_1,\overline{A}$ }; 
          \node[state, left of=l1] (l4) {$Q_4,\overline{A}$ }; 
          \node[state, right of=l2] (l3) {$Q_3,\overline{A}$ };

          \node[state, below of=l0] (l0c) {$Q_0,\overline{B}$ }; 
          \node[state, below of=l1] (l1c) {$Q_1,\overline{B}$ }; 
          \node[state, below of=l4] (l4c) {$Q_4,\overline{B}$ }; 
          \node[state, right of=l0c] (l2c) {$Q_2,\overline{B}$}; 
          \node[state, below of=l3] (l3c) {$Q_3,\overline{B}$ };

          \node[state, below of=l0c] (l0d) {$Q_0,\overline{C}$ }; 
          \node[state, below of=l1c] (l1d) {$Q_1,\overline{C}$ }; 
          \node[state, below of=l4c] (l4d) {$Q_4,\overline{C}$ }; 
          \node[state, below of=l2c] (l2d) {$Q_2,\overline{C}$}; 
          \node[state, below of=l3c] (l3d) {$Q_3,\overline{C}$ };

          \path[->, font=\Large]
            (l0c) edge node[above] {$a$} (l1)
            (l0d) edge node[above] {$a$} (l1)
            (l0c) edge node[above]  {$a $} (l2c)
            (l2c) edge  node[above] {$b$} (l3)
            (l2d) edge  node[above] {$b$} (l3)
            (l1) edge node[above] {$\epsilon$} (l4)
            (l1c) edge node[above] {$\epsilon$} (l4c)
            (l4c) edge [bend left] node[right] {$b$} (l4)
            (l4d) edge [bend left] node[left] {$b$} (l4)
                (l3c) edge [bend right] node[left] {$b$} (l3)
                (l3d) edge [bend right] node[right] {$b$} (l3)
            (l0) edge node[right] {$\checkmark$} (l0c)
            (l0c) edge node[right] {$\checkmark$} (l0d)
            (l1) edge node[left] {$\checkmark$} (l1c)
            (l1c) edge node[left] {$\checkmark$} (l1d)
            (l4) edge node[right] {$\checkmark$} (l4c)
            (l4c) edge node[right] {$\checkmark$} (l4d)
            (l3) edge node[left] {$\checkmark$} (l3c)
            (l3c) edge node[left] {$\checkmark$} (l3d)
            (l2c) edge node[left] {$\checkmark$} (l2d)
            (l0d) edge[loop below] node {$\checkmark$} ()
            (l1d) edge[loop below] node {$\checkmark$} ()
            (l2d) edge[loop below] node {$\checkmark$} ()
            (l3d) edge[loop below] node {$\checkmark$} ()
            (l4d) edge[loop below] node {$\checkmark$} ()
            (l4) edge[loop above] node {$b$} ();
        \end{tikzpicture}
    }
    \caption{The integral automaton $\mathcal{B}^{\checkmark}$ in Example \ref{ex:2}, where
   ``$\overline{A}$",``$\overline{B}$", and ``$\overline{C}$" denote the regions of $[x=0]$, $[x=1]$, and $[x=2] (i.e., [x>1])$, respectively, and all unreachable states are erased.}
    \label{fig8}
\end{figure}

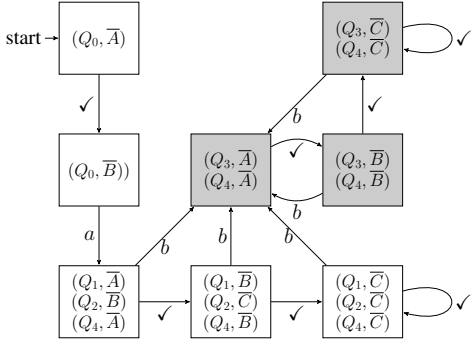
\begin{figure}[!h]
    \centering
    \resizebox{0.35\textwidth}{!}{%
        \begin{tikzpicture}[font=\huge,->,>=stealth',shorten >=1pt,auto,semithick, node distance=4.5cm,
                                  every state/.style={rectangle,minimum width=0.15\textwidth, minimum height=7em, font=\LARGE},
                                  new-qs/.style={fill=gray!40,thick}]
                \node[initial, state] (x0) {$(Q_0,\overline{A})$};
                \node[state]  [below of = x0] (x1) {$(Q_0,\overline{B}))$};
                \node[state, align=center]   [below of = x1] (x2) {$(Q_1,\overline{A})$ \\ $(Q_2,\overline{B}) $ \\ $(Q_4,\overline{A})$};
                 \node[state,new-qs, align=center]  [right of = x1] (x3) {$(Q_3,\overline{A})$ \\ $(Q_4,\overline{A})$};

      \node[state, align=center, right of = x2] (x4) {$(Q_1,\overline{B})$ \\ $(Q_2,\overline{C})$ \\ $(Q_4,\overline{B})$};    \node[state, align=center, right of= x4] (x5) {$(Q_1,\overline{C})$  \\ $(Q_2,\overline{C})$ \\ $(Q_4,\overline{C})$};
      \node[state,new-qs, align=center, right of = x3] (x6) {$(Q_3,\overline{B})$ \\$(Q_4,\overline{B})$ };
      \node[state,new-qs, align=center, above of = x6] (x7) {$(Q_3,\overline{C})$ \\$(Q_4,\overline{C})$ };

                \path[->]
                (x0) edge node[left]  {$\checkmark$} (x1)
                (x1) edge node[left]  {$a$} (x2)
                (x2) edge node[below]  {$\checkmark $} (x4)
                (x4) edge node[below]  {$\checkmark $} (x5)
                (x6) edge node[right]  {$\checkmark $} (x7)
                (x3) edge [bend left ]node[below]  {$\checkmark $} (x6)
                (x6) edge [bend left ]node[below]  {$b $} (x3)
                (x7) edge node[below]  {$b $} (x3)
                (x2) edge node[below]  {$b $} (x3)
                (x4) edge node[left]  {$b $} (x3)
                (x5) edge node[left]  {$b $} (x3)
                 (x5) edge[loop right] node {$\checkmark$} ()
                 (x7) edge[loop right] node {$\checkmark$} ();
                ;
        \end{tikzpicture}
    }
     \caption{A part of $\mathscr{H}$ in Example \ref{ex:2},
     where only the reachable states $X$ satisfying $\overline{L}_{X} \cap L \neq \emptyset$
     or the states that can reach those states $X$ are shown. }
    \label{fig9}
\end{figure}

\end{example}

\section{Conclusion}

This paper has established the necessary and sufficient conditions
under which location-based opacity becomes decidable in TA,
 addressing an open problem posed by An \emph{et al.} (FM2024).
 Leveraging these conditions, we have identified IRTA as a novel subclass of TA where opacity verification is decidable,
 and we have proposed a verification algorithm for CLTO in IRTA.
 Moreover, by assuming intruders observe time only in discrete units,
 we have introduced a relaxed opacity notion, termed CLTO-IDTP, and also
  provided a verification algorithm for CLTO-IDTP.
 These contributions establish theoretical foundations for modeling timed systems and intruders in security analysis,
 enabling the construction of timed models that effectively balance expressiveness and decidability.
 \par

 However, our current opacity verification algorithms depend on region automaton construction,
 resulting in high time complexity.
 Future work will focus on developing more efficient verification techniques,
 such as symbolic representations or abstraction refinement,
 and exploring practical applications of IRTA and CLTO-IDTP in cybersecurity for real-time systems.

\section*{Acknowledgments}
This work is supported by the National Natural Science Foundation of China (Grant No. 61876195),
 the Natural Science Foundation of Guangdong Province of China (Grant Nos. 2022A1515011136, 2025A1515012808),
  Guangxi Science and Technology Project (No. Guike AD23026227)
  and the Project Improving the Basic Scientific Research Ability of Young and Middle-aged Teachers
  in Guangxi Universities of China (Grant No. 2021KY0591).


\begin{thebibliography}{1}
\bibliographystyle{IEEEtran}

\bibitem{desbook}
C. G. Cassandras and S. Lafortune, {\it{Introduction to Discrete Event Systems}}, 2nd Ed.,
 New York, NY, USA: Springer, 2008.

\bibitem{opacity-review}
 R. Jacob, J. J. Lesage, and J. M. Faure, ``Overview of discrete event systems opacity:
 models, validation, and quantification," \emph{Annual Reviews in Control}, vol. 41, pp. 135-146, 2016.

\bibitem{l-opacity}
F. Lin, ``Opacity of discrete event systems and its applications,"
\emph{Automatica}, vol. 47, no. 3, pp. 496-503, 2011.

\bibitem{cso}
A. Saboori and C. N. Hadjicostis, ``Notions of security and opacity in discrete event systems,"
in \emph{Proceedings of the 46th IEEE Conference on Decision and Control (CDC 2007)}, New Orleans, LA, USA, 2007,
pp. 5056-5061.

\bibitem{ifo}
Y. Wu and S. Lafortune, ``Comparative analysis of related notions of opacity in centralized and coordinated architectures," \emph{Discrete Event Dynamic Systems}, vol. 23, no. 3, pp. 307-339, 2013.

\bibitem{infinite}
 X. Yin and S. Lafortune, ``A new approach for the verification of infinite-step and K-step opacity using two-way observers," \emph{Automatica}, vol. 80, pp. 162-171, 2017.

\bibitem{pre-opacity}
S. Yang and X. Yin, ``Secure your intention: on notions of pre-opacity in discrete-event systems,"
\emph{IEEE Transactions on Automatic Control}, vol. 68, no. 8, pp. 4754-4766, 2023.

\bibitem{probilistic-opacity1}
 A. Saboori and C. N. Hadjicostis, ``Current-state opacity formulations in probabilistic finite automata,"
  \emph{IEEE Transactions on Automatic Control}, vol. 59, no. 1, pp. 120-133, 2014.

\bibitem{fuzzy-opacity2}
W. Deng, D. Qiu, and J. Yang, ``Fuzzy infinite-step opacity measure of discrete event systems and its applications,"
\emph{IEEE Transactions on Fuzzy Systems}, vol. 30, no. 3, pp. 885-892, 2022.

\bibitem{petri-opacity1}
Y. Tong, Z. Li, C. Seatzu, and A. Giua, ``Verification of state-based opacity using Petri nets,"
\emph{IEEE Transactions on Automatic Control}, vol. 62, no. 6, pp. 2823-2837, 2017.


\bibitem{ad94-timed-automata}
 R. Alur and D. L. Dill, ``A theory of timed automata,"
  \emph{Theoretical Computer Science}, vol. 126, no. 2, pp. 183-235, 1994.



\bibitem{timed-opacity}
F. Cassez, ``The dark side of timed opacity," in \emph{Proceedings of the 3rd International Conference on Information
Security and Assurance (ISA 2009)}, Seoul, South Korea, 2009, pp. 21-30.


\bibitem{era1}
 R. Alur, L. Fix, and T. A. Henzinger, ``Event-clock automata: a determinizable class of timed automata,"
  \emph{Theoretical Computer Science}, vol. 211, no. 1, pp. 253-273, 1999.


\bibitem{wang2018-realtime-opacity}
L. Wang, N. Zhan, and J. An, ``The opacity of real-time automata,"
 \emph{IEEE Transactions on Computer-Aided Design of Integrated Circuits and Systems},
 vol. 37, no. 11, pp. 2845-2856, 2018.

\bibitem{Zhang2021-realtime-opacity}
K. Zhang, ``State-based opacity of real-time automata," in \emph{Proceedings of the 27th International Workshop on Cellular Automata and Discrete Complex Systems (AUTOMATA 2021)}, Marseille, France, 2021, Article No. 12, pp. 1-15.

\bibitem{zhang2024-realtime-opacity}
K. Zhang, ``State-based opacity of labeled real-time automata,"  \emph{Theoretical Computer Science},
vol. 987, Article No. 114373, pp. 1-20, 2024.

\bibitem{li2022-realtime-opacity}
J. Li, D. Lefebvre, C. N. Hadjicostis, and Z. Li, ``Observers for a Class of Timed Automata Based on Elapsed Time Graphs,"
\emph{IEEE Transactions on Automatic Control}, vol. 67, no. 2, pp. 767-779, 2022.

\bibitem{li2025-realtime-opacity}
J. Li, D. Lefebvre, C. N. Hadjicostis, and Z. Li, ``Verification of state-based timed opacity for constant-time labeled automata," \emph{IEEE Transactions on Automatic Control}, vol. 70, no. 1, pp. 503-509, 2025.

\bibitem{An2024}
J. An, Q. Gao, L. Wang, N. Zhan, and I. Hasuo, ``The opacity of timed automata,"
in \emph{Proceedings of the 26th International Symposium on Formal Methods (FM 2024)}, Milan, Italy, 2024, pp. 620-637.

\bibitem{Andre2024}
\'{E}. Andr\'{e}, S. D\'{e}pernet, and E. Lefaucheux, ``The bright side of timed opacity,"
in \emph{Proceedings of the 25th International Conference on Formal Engineering Methods (ICFEM 2024)},
Hiroshima, Japan, 2024, pp. 51-69.

\bibitem{Klein2024}
J. Klein, P. Kogel, and S. Glesner, ``Verifying opacity of discrete-timed automata,"
 in \emph{Proceedings of IEEE/ACM 12th International Conference on Formal Methods in Software Engineering (FormaliSE 2024)}, Lisbon, Portugal, 2024, pp. 55-65.


\bibitem{Andre-execution-opacity}
\'{E}. Andr\'{e}, M. Duflot, L. Laversa, and E. Lefaucheux, ``Execution-time opacity control for timed automata,"
in \emph{Proceedings of the 22nd International Conference on Software Engineering and Formal Methods (SEFM 2024)},
Aveiro, Portugal, 2024, pp. 347-365.

\bibitem{Ammar-bounded-opacity}
I. Ammar, Y. El Touati, M. Yeddes, and J. Mullins, ``Bounded opacity for timed systems,"
\emph{Journal of Information Security and Applications}, vol. 61, Article No. 102926, pp. 1-13, 2021.

\bibitem{IRTA}
P. V. Suman, P. K. Pandya, S. N. Krishna, and L. Manasa, ``Timed Automata with Integer Resets: Language Inclusion and Expressiveness," in \emph{Proceedings of the 6th International Conference on Formal Modeling and Analysis of Timed Systems},
Saint-Malo, France, 2008, pp. 78-92.


\bibitem{digitization1}
T. A. Henzinger, Z. Manna, and A. Pnueli, ``What good are digital clocks?"
in \emph{Proceedings of the 19th International Colloquium on Languages and Programming (ICALP 1992)},
Wien, Austria, 1992, pp. 545-558.

\bibitem{digitization2}
J. Ouaknine and J. Worrell, ``Revisiting digitization, robustness, and decidability for timed automata,"
 in \emph{Proceedings of the 18th Annual IEEE Symposium of Logic in Computer Science (LICS 2003)}, Ottawa,
  ON, Canada, 2003, pp. 198-207.

\bibitem{digitization3}
T. A. Henzinger and J.-F. Raskin, ``Robust undecidability of timed and hybrid systems,"
 in \emph{Proceedings of the 3rd International Workshop on Hybrid Systems: Computation and Control (HSCC 2000)},
 Pittsburgh, PA, USA, 2000, pp. 145-159.

 \bibitem{simulation}
L. Ilie, G. Navarro, and S. Yu, ``On NFA reductions," in \emph{Theory is Forever, Lecture Notes in Computer Science}, vol. 3113, Berlin, Heidelberg: Springer, 2004, pp. 112-124.
\end{thebibliography}
\end{document}